\documentclass[
aip, pop,
 amsmath,amssymb,
 reprint,%
tightenlines,
flushbottom,
citeautoscript,
floatfix
]{revtex4-1}

\usepackage{graphicx}% Include figure files
\usepackage{dcolumn}% Align table columns on decimal point
\usepackage{bm}% bold math
\usepackage{comment}
\usepackage{booktabs}
\usepackage{wasysym}
\usepackage[separate-uncertainty=true]{siunitx}
\usepackage[utf8]{inputenc}
\usepackage[T1]{fontenc}
\usepackage{mathptmx}
\usepackage{etoolbox}
\usepackage{hyperref}
\hypersetup{
    colorlinks = true,
    linkcolor = blue,
    citecolor = blue,
    urlcolor=blue,
}
\usepackage{tabularx}
\usepackage{xcolor}
\usepackage{xspace}

\DeclareMathOperator\erf{erf}
\newcommand{\neL}{\(\langle n_e L\rangle\)\xspace}

\makeatletter
\def\@email#1#2{%
 \endgroup
 \patchcmd{\titleblock@produce}
  {\frontmatter@RRAPformat}
  {\frontmatter@RRAPformat{\produce@RRAP{*#1\href{mailto:#2}{#2}}}\frontmatter@RRAPformat}
  {}{}
}%
\makeatother

\usepackage{setspace}
\usepackage{parskip}
\usepackage[compact]{titlesec}         % you need this package
\titlespacing{\section}{4pt}{4pt}{4pt} % this reduces space between (sub)sections to 0pt, for example
\AtBeginDocument{%                     % this will reduce spaces between parts (above and below) of texts within a (sub)section to 0pt, for example - like between an 'eqnarray' and text
  \setlength\abovedisplayskip{4pt}
  \setlength\belowdisplayskip{4pt}
  }

\usepackage{natmove}
\begin{document}

% \title{Towards a Pulsed-Power-Driven Platform For Studying Magnetic Reconnection With A Guide Field} 
\title{Quadrupolar Density Structures in Driven Magnetic Reconnection Experiments with a Guide Field}



\author{T. W. O. Varnish}
\affiliation{Plasma Science and Fusion Center, Massachusetts Institute of Technology, MA 02139, Cambridge, USA\looseness=-10000 %\\This line break forced with \textbackslash\textbackslash
}%

\author{J. Chen}
\affiliation{Department of Nuclear Engineering and Radiological Sciences, University of Michigan, Ann Arbor, MI 48109, USA\looseness=-10000 %\\This line break forced with \textbackslash\textbackslash}
}
\author{S. Chowdhry}
\affiliation{Plasma Science and Fusion Center, Massachusetts Institute of Technology, MA 02139, Cambridge, USA\looseness=-10000 %\\This line break forced with \textbackslash\textbackslash
}%

\author{R. Datta}
\affiliation{Plasma Science and Fusion Center, Massachusetts Institute of Technology, MA 02139, Cambridge, USA\looseness=-10000 %\\This line break forced with \textbackslash\textbackslash
}%

\author{G. V. Dowhan}
\altaffiliation[Current address: ]{Naval Research Laboratory, Washington, DC 20375, USA}
\affiliation{Department of Nuclear Engineering and Radiological Sciences, University of Michigan, Ann Arbor, MI 48109, USA\looseness=-10000 %\\This line break forced with \textbackslash\textbackslash}
}
\author{L. S. Horan IV}
\affiliation{Plasma Science and Fusion Center, Massachusetts Institute of Technology, MA 02139, Cambridge, USA\looseness=-10000 %\\This line break forced with \textbackslash\textbackslash
}%

\author{N. M. Jordan}
\affiliation{Department of Nuclear Engineering and Radiological Sciences, University of Michigan, Ann Arbor, MI 48109, USA\looseness=-10000 %\\This line break forced with \textbackslash\textbackslash}
}
\author{E. R. Neill}
\affiliation{Plasma Science and Fusion Center, Massachusetts Institute of Technology, MA 02139, Cambridge, USA\looseness=-10000 %\\This line break forced with \textbackslash\textbackslash
}%

\author{A. P. Shah}
\altaffiliation[Current address: ]{University of California---San Diego, La Jolla, CA 92093}
\affiliation{Department of Nuclear Engineering and Radiological Sciences, University of Michigan, Ann Arbor, MI 48109, USA\looseness=-10000 %\\This line break forced with \textbackslash\textbackslash}
}
\author{B. J. Sporer}
\altaffiliation[Current address: ]{TAE Technologies, Inc., Irvine, CA, USA}
\affiliation{Department of Nuclear Engineering and Radiological Sciences, University of Michigan, Ann Arbor, MI 48109, USA\looseness=-10000 %\\This line break forced with \textbackslash\textbackslash}
}
\author{R. Shapovalov}
\altaffiliation[Current address: ]{Laboratory for Laser Energetics, Rochester, NY, USA}
\affiliation{Department of Nuclear Engineering and Radiological Sciences, University of Michigan, Ann Arbor, MI 48109, USA\looseness=-10000 %\\This line break forced with \textbackslash\textbackslash}
}
\author{R. D. McBride}
\affiliation{Department of Nuclear Engineering and Radiological Sciences, University of Michigan, Ann Arbor, MI 48109, USA\looseness=-10000 %\\This line break forced with \textbackslash\textbackslash}
}
\author{J. D. Hare}
\email[]{jdhare@cornell.edu}
\affiliation{Plasma Science and Fusion Center, Massachusetts Institute of Technology, MA 02139, Cambridge, USA\looseness=-10000 %\\This line break forced with \textbackslash\textbackslash
}%
\affiliation{Laboratory of Plasma Studies, Cornell University, Ithaca, NY 14853, USA\looseness=-10000 %\\This line break forced with \textbackslash\textbackslash
}%

\date{\today}

\begin{abstract}

Magnetic reconnection is a ubiquitous process in plasma physics, driving rapid and energetic events such as coronal mass ejections.
Reconnection between magnetic fields with arbitrary shear can be decomposed into an anti-parallel, reconnecting component, and a non-reconnecting guide-field component which is parallel to the reconnecting electric field. 
This guide field modifies the structure of the reconnection layer and the reconnection rate.
We present results from experiments on the MAIZE pulsed-power generator (500 kA peak current, 200 ns rise-time) which use two exploding wire arrays, tilted in opposite directions, to embed a guide field in the plasma flows with a relative strength $b\equiv B_g/B_{rec}=0$, 0.4, or 1. 
The reconnection layers in these experiments have widths which are less than the ion skin depth, $d_i=c/\omega_{pi}$, indicating the importance of the Hall term, which generates a distinctive quadrupolar magnetic field structure along the separatrices of the reconnection layer.
Using laser imaging interferometry, we observe quadrupolar structures in the line-integrated electron density, consistent with the interaction of the embedded guide field with the quadrupolar Hall field.
Our measurements extend over much larger length scales ($40 d_i$) at higher $\beta$ ($\sim 1$) than previous experiments, providing an insight into the global structure of the reconnection layer.
\end{abstract}

\maketitle 
\section{\label{sec:intro} Introduction}

Magnetic reconnection is an explosive process in which magnetic field lines rapidly change their topology within a narrow region known as the reconnection layer.
During this process, magnetic energy is converted into the kinetic energy of fast reconnection outflows from the layer, and into the thermal and non-thermal energy of the particles inside the layer.\cite{Zweibel2009, Zweibel2016}

In many plasmas, the reconnecting magnetic field lines are not exactly anti-parallel, and the magnetic field can be separated into two orthogonal components: an anti-parallel component, \(B_{rec}\), which lies in the plane of reconnection, and a guide field component, \(B_g\), which is normal to this plane and hence parallel to the reconnecting electric field.
This guide field contributes to the pressure balance within the reconnection layer, which alters the layer morphology and reconnection rate,\cite{Huba2004, Yang2006} as well as modifying the partition of energy,\cite{Cassak2017} and the mechanisms for particle acceleration.\cite{Egedal2012}
As such, guide field reconnection is important in a wide range of astrophysical and space plasmas, including in the solar wind\cite{Phan2010} and in the Earth's magnetotail \cite{Eastwood2010}.

In magnetic reconnection, two-fluid effects occur when the reconnection layer width ($\delta$) is on the order of the ion kinetic scales ($d_i = c/\omega_{pi}$ or $\rho_s = \Omega_i/c_s$).
On these scales, the electron and ion fluids decouple, resulting in a net in-plane electrical current which leads to a distinctive quadrupolar out-of-plane magnetic field,\cite{Uzdensky2006a} known as the Hall field.
This Hall field has been seen in simulations,\cite{Drake1992a,Huba2004} laboratory experiments,\cite{Ren2005a, Yamada2006} and in-situ spacecraft observations.\cite{Mozer2002, Wygant2005}
The Hall field is localised to the separatrices, and the additional magnetic pressure it provides causes a depletion of the electron density outside of the layer, which has been observed in simulations\cite{Huba2004, Yang2006, Yang2008} and in experiments.\cite{Hare2017c}

Two-fluid effects interact with a guide field to direct the outflows from the reconnection layer along one of the separatrices.
This forms a region of enhanced density along one separatrix, and a region of decreased density along the other separatrix, resulting in a quadrupolar pattern in the electron density, changing the symmetry of the reconnection layer.
This effect can clearly be seen in simulations, for example Fig. 3 of Yang \textit{et al.}\cite{Yang2008}, which has a guide field ratio $b = 1$.
A phenomenological explanation for this density distribution was given by Kleva \textit{et al.} \cite{Kleva1995}
Understanding these density structures on global scales is particularly important for interpreting spacecraft data, which represent one-dimensional cuts through complex three-dimensional structures.\cite{Eriksson2016}

Laboratory experiments have studied reconnection with a wide range of guide field ratios (\(b\equiv B_g/B_{rec}\)), and on current sheet widths from magneto-hydrodynamic (MHD) scales down to the ion and electron kinetic scales.
A recent review by Ji \textit{et al.}\cite{Ji2023pub} summarises many of these experiments, and so for the purposes of this paper we reference only those relevant to our experimental conditions, in which both a moderate guide field (\(b \sim 1\)) and ion kinetic scales ($\delta\leq d_i$ or $\rho_s$) are present.
In experiments on the CS-3D facility, Frank \textit{et al.}\cite{Frank2006} and Bogdanov \textit{et al.}\cite{Bogdanov2007} used holographic interferometry to observe a reconnection layer which appeared to be rotated, with a rotation angle which increased with guide field ratio.
We note that their explanation for the observed rotation of the reconnection layer is subtly different from that given by Yang \textit{et al.}\cite{Yang2008} and Kleva \textit{et al.}\cite{Kleva1995}.
They invoke a torque exerted on the reconnection layer plasma by the combination of the quadrupolar Hall and guide fields, rather than a redirection of the reconnection outflows.
This interpretation implies a time evolution of the layer rotation and does not predict a quadrupolar density structure, both of which are consistent with their results.
On the MRX facility, Tharp \textit{et al.}\cite{Tharp2012} and Fox \textit{et al.}\cite{Fox2017} measured the magnetic fields and electron pressure using in-situ probes, and showed a significant reduction in the reconnection rate with increasing guide field.
They also observed a quadrupolar variation in the out-of-plane magnetic field and the electron pressure---consistent with the model of Kleva \textit{et al.}\cite{Kleva1995}---and the importance of electron pressure gradients in an extended Ohm's Law.

Previous pulsed-power driven magnetic reconnection experiments have studied only anti-parallel reconnection, with no guide field.
These experiments used two exploding (or ``inverse'') wire arrays to produce cylindrically diverging sources of magnetised plasma---by driving two such arrays in parallel, the plasma flows collide with anti-parallel magnetic fields, and a reconnection layer forms between the arrays.\cite{Hare2018a}
In these experiments, the elemental composition of the plasma was found to have a significant effect on the reconnection process---with a higher-Z aluminium plasma, the super-Alfvénic flows led to shock compression of the layer at relatively a low Lundquist number (\(S\approx 10\))\cite{Suttle2016, Suttle2018}, whereas in a lower-Z carbon plasma we observed plasmoids formed by the tearing instability at a higher Lundquist number \(S\approx 100\).\cite{Hare2017, Hare2017c}
In all of these experiments, these plasmas are collisional in the sense that the collisional mean-free-path is much less than the other length scales in the plasma, such as the electron and ion skin depths, and the width of the current sheet.
However, there was preliminary evidence for the existence of two-fluid reconnection (often called collisionless or semi-collisional reconnection) in the experiments with carbon, in which a density depletion region was observed outside of the reconnection layer,\cite{Hare2017c} consistent with predictions from simulations.\cite{Shay2001}
We emphasise here that an extended MHD Ohm's law allows both resistive (collisional) effects and the Hall term to be important in a plasma, and hence the presence of two-fluid effects is not a sufficient condition to describe the reconnection process as collisionless.

In this paper, we present results from  pulsed-power-driven magnetic reconnection experiments in which both the guide field and two-fluid effects are important.
We observed the formation of a distinct quadrupolar density distribution in a twisted current sheet, consistent with the redirection of the reconnection outflows predicted by theory and simulations.
We achieved this by tilting the two wire arrays in opposite directions, which directly embedded a guide field in the plasma flows.
The guide field strength ($b$) therefore depended on the tilt angle ($\theta$) as $b = \tan\theta$, and we carried out experiments with $b = $ 0, 0.4 and 1.
These tilted arrays created a twisted three-dimensional reconnection layer which is challenging to interpret using line-integrated diagnostics.
With a series of simple geometric models, as well as 3D magnetohydrodynamic simulations, we show that this observed density distribution cannot be explained by resistive magnetohydrodynamics alone.
However, when we consider the interaction between two-fluid effects and the guide field in our model, we are able to qualitatively reproduce our experimental results.
In contrast to previous experiments on CS-3D and MRX, we observed this quadrupolar density structure over larger length scales ($\approx 40 d_i$), providing insight into the global structure of a reconnection layer with two-fluid effects and a guide field.

\section{Methods} \label{sec:ta:method}
\begin{figure}[!h]
    \centering
    \includegraphics{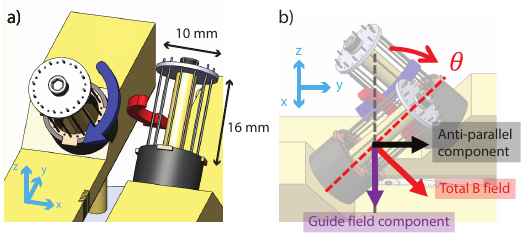}
    \caption{The tilted wire array geometry used in these experiments. a) A 3D model of the tilted geometry, showing that the two wire arrays are inclined in the $y-z$ plane. b) A view from the side, demonstrating that tilting the arrays by an angle $\theta$ results in an azimuthal magnetic field around each array that can be decomposed into orthogonal anti-parallel and guide field components.}
    \label{fig:ta_setup}
\end{figure}

We designed a scaled-down version of the reconnection platform developed for MAGPIE (\SI{1.4}{\mega\ampere} peak current, \SI{240}{\nano\second} rise time),\cite{Hare2018a} suitable for the lower peak current on MAIZE (in this experimental series, \SI{500}{\kilo\ampere} peak current, \SI{200}{\nano\second} rise time).
Fig. \ref{fig:ta_setup} shows this configuration: each wire array was \SI{10}{\mm} in diameter and \SI{16}{\mm} tall, with eight \SI{0.4}{\mm} diameter carbon rods (Staedlar Mars Micro Carbon B) spaced evenly around a \SI{4}{\mm} diameter central cathode.
The arrays had a centre-to-centre separation of \SI{22}{\mm}, giving a distance of \SI{6}{\mm} from the closest wires to the reconnection layer.

Unlike previous exploding wire array reconnection experiments, these arrays were tilted by an angle ($\theta$) to the vertical in opposite directions, as illustrated in Fig. \ref{fig:ta_setup}.
This reorients the azimuthal magnetic fields around each array, introducing a component of the field that points in the $z$ direction (out of the plane of reconnection): the guide field.
We note that this tilted setup is reminiscent of the laser-driven reconnection experiments of Bolaños \textit{et al.}\cite{Bolanos2022}
From the geometry, the tilt angle $\theta$ is related to the guide field ratio $b$ as follows:
\begin{equation}
    b = \frac{B_{g}}{B_{rec}} = \tan\theta.
\end{equation}
We fielded three sets of hardware, with tilt angles of \SI{0}{\degree}, \SI{22.5}{\degree}, and \SI{45}{\degree}, corresponding to guide field ratios of 0, 0.4, and 1, respectively.
The total magnetic field in the flows is independent of the tilt angle, $B_{tot} = (B_g^2+B_{rec}^2)^{1/2}$, and so reconnecting magnetic field decreases with increasing tilt angle (guide field).

We diagnosed these experiments using laser interferometry along two orthogonal lines-of-sight, a 12-frame optical camera, a four-frame XUV camera, and ``B-dot'' inductive probes.
Results from the 12-frame optical camera and four-frame XUV camera were useful for checking the even distribution of current between the two wire arrays, but are not shown in this paper.
The laser interferometry used a \SI{50}{\micro\joule}, \SI{1064}{\nm}, \SI{2}{\ns} pulse (EKSPLA NL-122), with the beam expanded to a diameter of \SI{25}{\mm}. For each interferometry line-of-sight, we used a Mach-Zehnder configurations in which the probe beam passes through the plasma, and a reference beam passes around the plasma.
These two beams recombine on a beamsplitter and are imaged through the same optics onto a DSLR camera (Canon EOS DIGITAL REBEL XS, $3888\times2592$ pixels) with a long exposure (\SI{1}{\s}---the laser pulse-length sets the actual exposure time).
By tilting the recombining beam-splitter, we introduce a linear phase shift across the vacuum interferogram, which serves as the carrier signal which heterodynes the phase introduced by the plasma. 
We unfold these interferograms using the MAGIC2 code \cite{Swadling2014a, Hare2019} to produce line-integrated electron-density maps.
Interferograms were captured simultaneously using two Mach-Zehnder interferometers, one in a side-on configuration looking along the $y$-axis, and another in an end-on configuration looking along the $z$ axis.

Additionally, we measured the $B_{rec}$ magnetic field component in the flows using inductive (B-dot) probes.
These probes consist of two co-located loops (\SI{0.9}{\mm} loop diameter, AWG36 enamel-coated wire) with opposite polarity, and the two signals were digitized separately. 
This allows for common mode rejection, where the two signals (L and R)
\begin{align}
    V_L(t) &= -A_L\dot{B} + V_s(t), \\
    V_R(t) &= A_R\dot{B} + V_s(t)
\end{align}
can be subtracted from one another to recover the time-varying magnetic field component, $\dot{B}$, and remove any stray electrostatic signals. \cite{Hare2017c, Datta2022a}
The calibration coefficients $A_R$ and $A_L$ (the effective areas of the two probe loops) were determined independently before the shot using a calibration pulser.

\section{Results}

\subsection{B-dot probes}

\begin{figure}[!ht]
    \centering
    \includegraphics{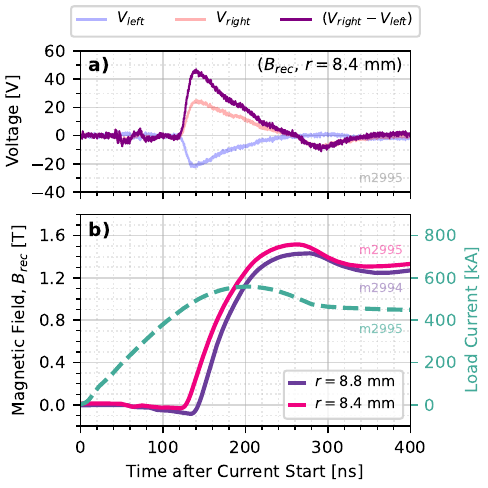}
    \caption{B-dots were positioned in the array outflows to measure the magnetic field advected by the plasma. (a) Raw voltage signals from the two counter-wound loops, measuring the reconnecting component of the magnetic field ($B_{rec}$). (b) Integrated differential signals from B-dots on two separate shots of the upright, $\SI{0}{\degree}$, $b=0$ hardware. Plotted on a separate axis is the current delivered to the experimental load for one of these shots.}
    \label{fig:ta_bdot}
\end{figure}

In Figure \ref{fig:ta_bdot}, we show results from a shot with a tilt angle of \SI{0}{\degree}, with the B-dot probes positioned at \SI{8.5\pm 0.5}{\mm} from the wires.
The raw signals for the probe (voltage proportional to $dB_{rec}/dt$) are shown in Fig. \ref{fig:ta_bdot}a, with the red and blue traces corresponding to the two oppositely wound loops, and common-mode rejection produces the purple signal which is proportional to $dB/dt$.
Fig. \ref{fig:ta_bdot}b shows the integrated magnetic fields for the probe in two separate shots, demonstrating the repeatability of our setup.
In both shots, the magnetic field starts to rise \SI{120}{\ns} after the current pulse, which reflects the time-of-flight of the plasma from the wires to the probe, corresponding to a flow velocity of \SI{70}{\km\per\s}.
This flow velocity is consistent with Thomson scattering measurements from previous pulsed-power-driven exploding wire array experiments \cite{Hare2017}.
The magnetic field rises more rapidly than the driving current, reaching a peak of around \SI{1.5}{\tesla} at \SI{260}{\ns}, consistent with results from MAGPIE (scaling for peak current and array radius).
Our measurements of the reconnecting magnetic field in the $b=0$ case also represent the total magnetic field expected in the $b=0.4$ and $b=1$ cases presented later, which in these tilted cases is divided between the reconnecting and guide field components.

\subsection{Interferometry}
\begin{figure}[!b]
    \centering
    \includegraphics{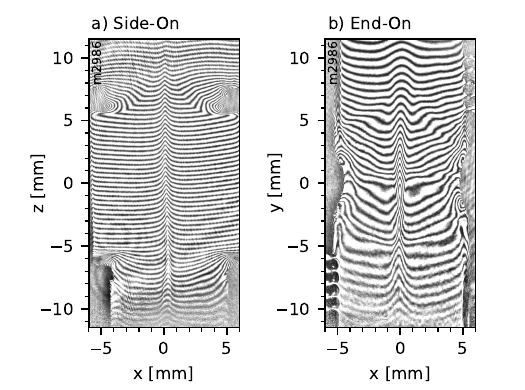}
    \caption{Two interferograms along orthogonal line-of-sight, taken during the same experiment ($b = 0$), at the same time: 
    a) looking along the $y$-axis (side-on view), and b) along the $z$-axis (end-on view).}
    \label{fig:ta_interferograms}
\end{figure}

\begin{figure*}[t]
    \includegraphics{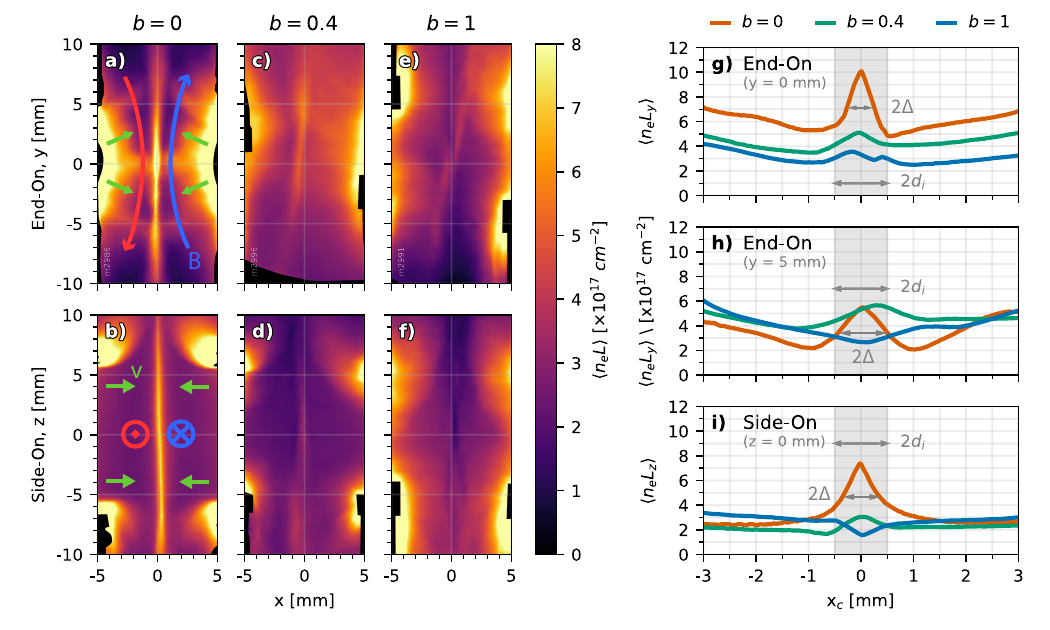}
    \caption{\label{fig:ta_results}
    (a--f) Line-integrated electron density maps taken at 320 ns after current start, in the end-on (a, c, e), and side-on (b, d, f) lines of sight. 
    From left to right, the guide field becomes stronger relative to the reconnecting field, from $b=0$ (a, b), to $b=0.4$ (c, d), to $b=1$ (e, f). 
    Line-outs from these maps for all guide field strengths are shown in g) along $y=\SI{0}{\mm}$, h) along $y=\SI{5}{\mm}$, and i) along $z=\SI{0}{\mm}$.
    In all cases, the reconnection layer width (as seen in the line-outs) is less than or on the order of the ion skin depth, $d_i$.
    }
\end{figure*}

Raw interferograms from the two orthogonal interferometers are shown in Fig. \ref{fig:ta_interferograms}.
Each interferogram consists of light and dark fringes, corresponding to the constructive and destructive interference between the reference beam and the probe beam.
The distortion of these fringes is due to the phase shift along the probing path, and it is this distortion that we process to unfold the line-integrated electron density.
These interferograms show excellent fringe contrast across the entire region of interest, and contain fiducials in the form of shadows from the load hardware, which we use to assign a coordinate system with the origin at the geometric centre of the hardware.

Figure \ref{fig:ta_results} presents line-integrated electron density (\neL) maps from experiments with three tilt angles (and therefore three guide field ratios, $b = 0, 0.4, 1$), with the end-on \neL maps on the top row and the corresponding side-on \neL maps below.
All data were taken at the same time, \SI{320}{\ns} after current start. 
Peak current occurred at \SI{200}{\ns} after current start for these shots.
Maps of \neL produced from interferograms provide only the relative line-integrated electron density---in order to provide an absolute measurement, the interferograms must contain a region free of plasma, known as the zero fringe shift region. 
This region is present in the $b = 0$ end-on interferograms, but not in the $b = 0$ side-on interferogram, nor in either interferogram for the $b = 0.4$ and $b = 1$ cases.
In order to estimate the zero in all of these \neL maps, we carried out the procedure detailed in Appendix \ref{sec:zeroing}.

The first column of Fig. \ref{fig:ta_results} shows the $b = 0$ (zero tilt angle) case, in which the advected magnetic fields are exactly anti-parallel and there is no guide field.
We observe the formation of a narrow region of enhanced line-integrated electron density in the side-on \neL map (Fig. \ref{fig:ta_results}b), which we interpret as a reconnection layer, consistent with previous experiments,\cite{Hare2018a}.
In the end-on line-of-sight (Fig. \ref{fig:ta_results}a), the flows coming from the left and the right arrays are modulated due to the discrete ablation flows from each wire, but the resulting layer is uniform and extends over a distance of approximately $\SI{11}{\mm}$, defined as the FWHM of density profile along $x = \SI{0}{\mm}$.
This is roughly equal to the radius of curvature of the magnetic field lines at the mid-plane, consistent with previous experiments and simulations of pulsed-power driven reconnection\cite{Hare2017}.
We note that density perturbations within the reconnection layer have previously been attributed to the formation of plasmoids by a secondary tearing instability,\cite{Hare2017} but we have no direct evidence to support that conclusion in these experiments, as we lack magnetic diagnostics in the outflow region.\cite{Hare2017c}

The second column of Fig. \ref{fig:ta_results} shows results from an experiment with a tilt of \SI{22.5}{\degree}, $b = 0.4$.
Similar to the zero guide field case, both the end-on and side-on \neL maps show a narrow region of enhanced \neL along $x = \SI{0}{\mm}$, consistent with the collision of the oppositely directed flows from the arrays on the left and the right.
Note that as the arrays are tilted, and the diagnostic is line-integrated, we no longer see the discrete flows in the end-on field-of-view.
We observe significant changes to the \neL map even with this relatively weak guide field: firstly, the contrast between the region of enhanced \neL and the inflows is lower in the $b = 0.4$ case compared to the $b = 0$ case, and secondly, this enhanced \neL region appears to be rotated in the end-on plane (Fig. \ref{fig:ta_results}c), a result which we will discuss in detail below.

In the third column of Fig. \ref{fig:ta_results}, we present results from the largest guide field ratio ($b = 1$) in this experimental series.
There are some similarities with the weaker guide field cases---the line-integrated electron density is still highest near the wires of the arrays on the left and right, and \neL drops off as the flows radially diverge outwards from the arrays towards the mid-plane.
In the end-on line-of-sight (Fig. \ref{fig:ta_results}e), we still see an enhanced region of line-integrated electron density, corresponding to the position at which the flows collided. 
This region appears to be even more rotated than in the $b = 0.4$ case, and the \neL within the enhanced region is lower than in the weaker guide field cases.
However, we see that this is not simply a rotation, but a quadrupolar density structure, as we observe a region of reduced line-integrated electron density---relative to the background \neL from the inflows---mirroring the structure of the enhanced \neL region left/right about $x = \SI{0}{\mm}$.
Combined, these enhanced/reduced \neL regions form a quadrupolar pattern in the end-on \neL map, centred approximately around the origin.
The side-on line-of-sight (Fig. \ref{fig:ta_results}f) shows an even more striking difference---there is a very pronounced dip in the line-integrated electron density at $x = \SI{0}{\mm}$, along with some complex structure at $z = \pm\SI{5}{\mm}$.

To make quantitative comparisons, we show line-outs from the end-on line-integrated density maps along $y = \SI{0}{\mm}$ (Fig. \ref{fig:ta_results}g) and $y = \SI{5}{\mm}$ (Fig. \ref{fig:ta_results}h), and from the side-on \neL maps along $z = \SI{0}{\mm}$ (Fig. \ref{fig:ta_results}i).
The horizontal axis $x_c$ is shifted from the $x$ axis in Figs. \ref{fig:ta_results}a-f such that the \neL profile peak is centred on $x_c = \SI{0}{\mm}$ at $y = \SI{0}{\mm}$, which corrects for small imperfections in the load hardware.
The $b = 0$ case (orange lines) shows a peaked \neL profile with a full-width-half-maximum of \(2\Delta \approx \SI{1}{\mm}\), centred at $x_c = \SI{0}{\mm}$ at both $y = \SI{0}{\mm}$ and $y = \SI{5}{\mm}$, indicating that there is no apparent rotation to the reconnection layer.

For the $b = 0.4$ case (green lines), the peaks of the \neL profiles are less pronounced, and the region of enhanced \neL appears to be broader.
At $y = \SI{5}{\mm}$, the \neL peak is shifted to $x_c = \SI{0.25}{\mm}$, consistent with an apparent rotation of the reconnection layer visible in Fig. \ref{fig:ta_results}.
However, upon closer inspection, at $x_c = \SI{-1}{\mm}$ (and $y = \SI{5}{\mm}$), there is also a slight reduction in the \neL relative to the background (compared with $x_c = \SI{1.25}{\mm}$).
Though at low contrast, this is consistent with a quadrupolar density structure.
In the line-out from the side-on field-of-view shown in Fig. \ref{fig:ta_results}i, the \neL profile is not symmetric about $x_c = \SI{0}{\mm}$ and has a dip in \neL at $x_c = \SI{0.5}{\mm}$.

\begin{figure*}[t]
    \includegraphics[width = 1\textwidth]{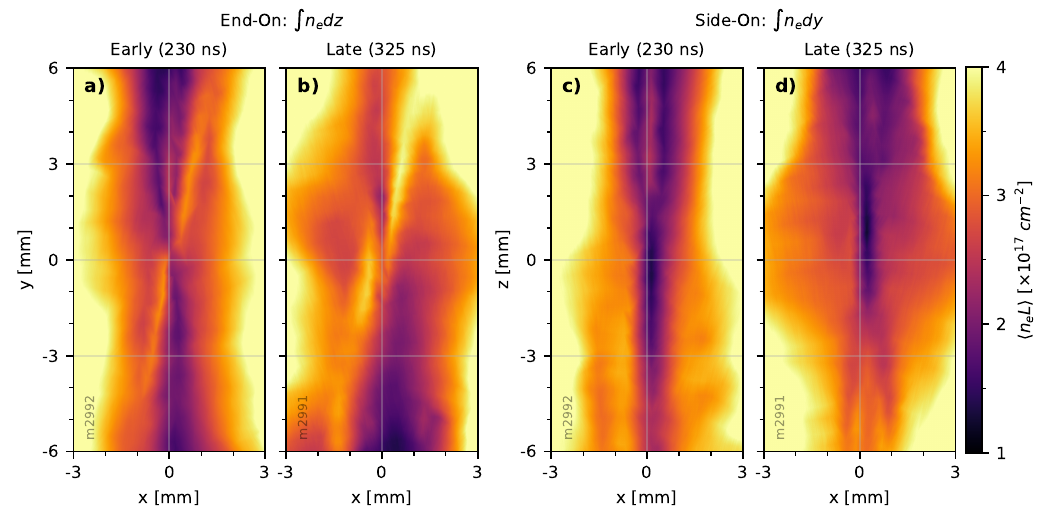}
    \caption{\label{fig:ta_results2}
    Line-integrated electron density maps from the end-on (a, b) and side-on (c,d) lines-of-sight, taken in two shots at different times:  \SI{230}{\ns} (a and c) and \SI{325}{\ns} (b and d), which were already shown in Fig. \ref{fig:ta_results}e and f.
    The figures are cropped around the centre of the layer, and the colourbars are adjusted to clearly show the quadrupolar structure visible in the end-on maps, and the up/down anti-symmetric structure in the side-on maps.
    }
\end{figure*}

For the strongest guide field, $b = 1$ (blue lines), the line-integrated electron density maps show complex structures.
In Fig. \ref{fig:ta_results}g ($y = \SI{0}{\mm}$), the \neL profile has a central peak, with a dip near $x_c = \SI{0.25}{\mm}$.
Similarly to the $b=0.4$ case, in the lineout at $y = \SI{5}{\mm}$ (Fig. \ref{fig:ta_results}h), the \neL maxima is displaced (to $x_c = \SI{1.5}{\mm}$), consistent with an apparent rotation of the layer.
We also see a minimum in the \neL lineout at $x_c = \SI{0}{\mm}$, consistent with the observed quadrupolar density structure.
In the lineout from the side on \neL map (Fig. \ref{fig:ta_results}i), there is a dip rather than a peak in \neL at $x_c = \SI{0}{\mm}$.

The striking line-integrated electron density features seen in the $b = 1$ case (Fig. \ref{fig:ta_results}e-f) are also seen at earlier times during the current pulse.
During a separate experiment, end-on and side-on interferograms were captured at $t = \SI{230}{\ns}$ after current start (compared to \SI{325}{\ns} for Fig. \ref{fig:ta_results}e-f), and these are shown in Fig. \ref{fig:ta_results2}a and c, along with the data from Fig. \ref{fig:ta_results}e and f, reproduced to aid direct comparison.

In the end-on \neL maps (Fig. \ref{fig:ta_results2}a and b), there is a clear quadrupolar structure relative to the background density, with a narrow region of increased \neL from the bottom left to the top right of the image, and a corresponding region of decreased \neL at roughly the same angle, running from the top left to the bottom right.
This structure is present at both early and late times, and the angle of these increased and decreased density regions has not changed significantly between these two shots, making it a reproducible and static feature over the observed time scales.

The side-on maps (Fig. \ref{fig:ta_results2}c and d) also exhibit complex structures, but along this line-of-sight, there is clear left/right symmetry.
In the bottom half of Fig. \ref{fig:ta_results2}c, there is a pronounced dip in electron density at $x = \SI{0}{\mm}$, with two regions of increased density flanking this dip at $x = \pm\SI{0.5}{\mm}$.
In the top half of the image, there is an increased region of line-integrated electron density along $x = \SI{0}{\mm}$, with two regions of decreased density flanking this dip at $x = \pm\SI{0.5}{\mm}$.
As such, there is also an antisymmetric up/down pattern to the \neL map, as well as the more obvious left/right symmetry.
The same features also appear in the late time map Fig. \ref{fig:ta_results2}d, showing again that these are reproducible features, and there is no significant time evolution over the observed period.

\section{Discussion}

Our results can be summarised by three main observations: 
\begin{enumerate}
    \item the line-integrated density near each wire array decreases with increasing guide field ratio $b$, 
    \item the line-integrated density in the reconnection layer decreases with increasing guide field ratio $b$, and 
    \item the layer observed in the end-on line-of-sight has an apparent rotation, by an angle which increases with the guide field. 
    Closer inspection reveals this is in fact a quadrupolar density structure with a region of increased density and a region of decreased density.
\end{enumerate}

The explanation for the first observation is trivial---the decrease in \neL near each wire array is expected even for a single exploding wire array, as the integration path length will change as the tilt angle is increased.
Some discussion of this effect is included in Appendix \ref{sec:zeroing}, which deals with the zeroing of the line-integrated electron density maps.

For the second and third observations---regarding the layer density and structure---the three-dimensional shape of the reconnection layer is clearly important.
In the \SI{0}{\degree} (\(b = 0\)) case, the reconnection layer is simply a planar region of increased electron density, which forms halfway between the two arrays.
For the tilted cases ($b = $ 0.4 and 1), however, the three-dimensional structure of the reconnection layer is not obvious.
Because we only have two orthogonal line-integrated measurements of the electron density, we cannot uniquely reconstruct the three-dimensional structure of the layer from our observations.

We first try to explain these observations using only geometric models for the shape of the reconnection layer, based on the tilt of the two exploding wire arrays. 
We begin with an argument from symmetry, and then extend this to a simple analytical model based on the rocket model, before considering three-dimensional magneto-hydrodynamic (MHD) simulations.
For each model, we see that the predicted three-dimensional geometry of the layer does not explain the quadrupolar density structures seen in the \neL maps in Fig. \ref{fig:ta_results}.
These structures must therefore be due to the physics inside the reconnection layer.
We then present evidence from two-fluid simulations of reconnection with a guide field which qualitatively matches the observed line-integrated densities.

\subsection{An analytical model for the reconnection layer shape}

We note that the array hardware has rotational symmetry around the $x$, $y$, and $z$ axes, which will be shared by the electron density distribution $n_e(x,y,z)$ produced by the two arrays.
As such, the line-integrated electron density maps must have mirror symmetries $[\int n_e dz] (x,y) = [\int n_e dz] (x,-y)$ etc, precluding any quadrupolar structures.
The full version of this symmetry argument is presented in Appendix \ref{sec:symmetry}.

This symmetry argument does not, however, predict the shape of the layer.
Instead, we predict the layer shape using the well-known rocket model for ablation from a wire array \cite{Lebedev2001}.
Motivated by the planar shape of the layer in the untilted ($b = 0$) case, we assume that the centroid of the layer lies where the densities in the flows from each array are equal, which from the rocket model is where the distance from the axis of each array is equal, $r_1 = r_2$.
Figure \ref{fig:ta_model} illustrates the setup of this model, and the full derivation is provided in Appendix \ref{app:geometric_model}. 

\begin{figure}
    \includegraphics[width = \linewidth]{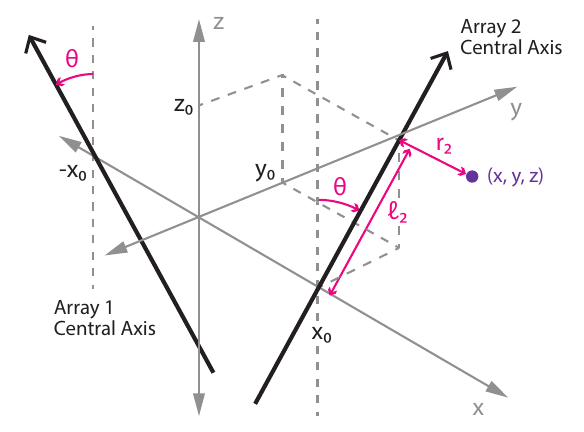}
    \caption{\label{fig:ta_model}
    Two arrays, 1 and 2, are centred around $-x_0$ and $x_0$, respectively.
    Each is tilted in opposite directions by an angle, $\theta$.
    The radial distance from the axis of array 2 to a point $(x, y, z)$ is $r_2$.
    }
\end{figure}

Our model gives the predicted geometric shape of the reconnection layer centroid, parametrised by $b$ and $x_0$ (half the separation of the two arrays):
\begin{equation}
    y = -\frac{x_0 (1 + b^2)}{b} \cdot \frac{1}{z} \cdot x.
\end{equation}
For $\theta = 0$ ($b=0$), this gives us a layer where $x = 0$ for all $y$, $z$, and so we recover the standard planar reconnection layer for the case of zero-guide field.
For any non-zero tilt angle, the layer in a given $z$ plane is a straight line, $y = mx$, and for a tilt angle of $\theta=\SI{45}{\degree}$ ($b=1$) we find
\begin{equation}\label{eqn:layer_angle}
    y = -\frac{2x_0}{z} \cdot x.
\end{equation}

\begin{figure}[!ht]
    \centering
    \includegraphics[width = \columnwidth]{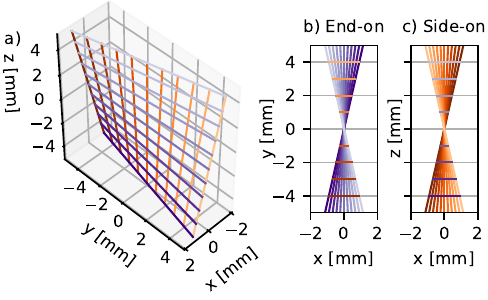}
    \caption{Model of the layer centroid from eqn. \ref{eqn:layer_angle}, which describes a hyperbolic paraboloid. The orange lines are at constant $y'$ in the $x{-}z$ plane, $x = -zy'/2x_0$, and are shaded from dark (negative $y'$) to light (positive $y'$). The purple lines lie in the $x{-}y$ plane, $x = -z'y/2x_0$, and are shaded from dark (negative $z'$) to light (positive $z'$). a) 3D perspective. b) End-on view. c) Side-on view.}
    \label{fig:hyper_para}
\end{figure}

For $z=0$, we find $x=0$ as expected for an untwisted layer, but for planes away from the $z = 0$ plane, we find straight-line slices of the layer which are increasingly rotated as $z$ increases.
Considered as an object in three dimensions, this is a doubly-ruled surface known as a hyperbolic paraboloid (see Fig. \ref{fig:hyper_para}) with the same symmetries as the load hardware (Appendix \ref{sec:symmetry}). 
Due to this symmetry, line integrations along $x$, $y$ or $z$ will produce line-integrated electron density maps with mirror symmetry around the two remaining axes.
Therefore, this simple analytical model predicts line-integrated electron density maps with strict symmetry, despite the prediction that the layer is a twisted, three-dimensional object.
This contradicts our experimental observations of a quadrupolar density structure, and as such we must consider physical effects---such as the Hall term---which allow the electron density to have different symmetries than the load hardware.

This model has several limitations: it only predicts where the centre of the layer is, rather than the width of the layer; it assumes axisymmetry of the flows from the two wire arrays, and so does not include the discrete nature of the wires; and it does not consider the actual interaction of the plasma flows, nor the acceleration of the outflows from the reconnection layer.

\subsection{Magneto-hydrodynamic simulations of the reconnection layer} \label{sec:ta:sims}

\begin{figure*}
    \includegraphics[width=1.0\textwidth]{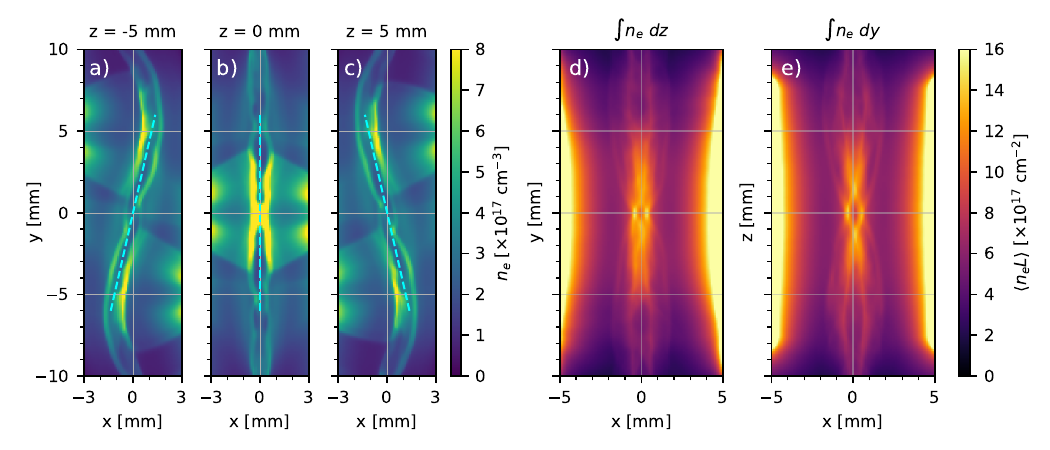}
    \caption{\label{fig:ta_sims}
    Results from simulations of tilted wire arrays using the GORGON 3D MHD code. The arrays are tilted at $\theta=\SI{45}{\degree}$ ($b=1$). Panels a)-c) show slices of $n_e$ in the $(x,y)$ plane at $z = $ -5, 0, and 5 mm, showing the geometric twist of the reconnection layer. The layer angle predicted by eqn. \ref{eqn:layer_angle} is overlaid as a cyan dashed line. Panels d) and e) show the line-integrated electron density (the experimental observable) in the end-on and side-on orientations, respectively. Despite the twist in the layer seen in panels a)-c), the line-integrated electron density maps show no overall rotation.
    }
\end{figure*}

We build on the simple analytical model discussed above using three-dimensional magneto-hydrodynamic simulations using the GORGON code, an Eulerian resistive MHD code with van Leer advection.\cite{Chittenden2004, Ciardi2007} 
We model the tilted arrays using tilted central cathodes, but we implement untilted (horizontal) anode and cathode disks to simplify the boundary conditions.
We use a \SI{100}{\um} grid with $480\times400\times280$ cells in $(x,y,z)$, and we drive the arrays using an azimuthal time-varying magnetic field boundary condition inside each coaxial transmission line for each array at the bottom $(z = \SI{-14}{\mm})$ of the simulation domain.
The magnetic field is set to reproduce a $I = \SI{530}{\kilo\ampere} \sin^2\left[(\pi/2) (t/\SI{200}{\ns})\right]$ current pulse representative of that produced by MAIZE during these experiments (and specifically matched to shot m2992).

We focus on the electron densities predicted by these simulations, as these are quantities for which we have experimental data in the form of the \neL maps shown in Fig. \ref{fig:ta_results}.
Slices of the electron density in the $(x,y)$ plane are shown in Fig. \ref{fig:ta_sims}a-c, at $z= $ -5, 0, and 5 mm, at \SI{200}{\nano\second}.
These slices show a twisted reconnection layer, in excellent agreement with the prediction of eqn. \ref{eqn:layer_angle}, despite the more complex structure imposed by the discrete wires which is not captured by the rocket model.

The synthetic line-integrated electron density maps are shown in Fig. \ref{fig:ta_sims}d-e, representing the numerical integration along the $z$ and $y$ axes respectively, as in the experimental results in Fig. \ref{fig:ta_results}.
Notably, despite the twist in the layer, the line-integrated electron density maps show no quadrupolar density feature in the electron density, a result which is consistent with our symmetry argument and geometric rocket model above.
As such, resistive MHD simulations do not accurately predict the quadrupolar density variation seen in the end-on electron density maps for the $b=1$ case in Fig. \ref{fig:ta_results2}, nor the up/down anti-symmetry seen in the side-on maps.

We note that the simulated densities are significantly higher than observed in the experiments, by a factor of around two to three.
We currently do not have an explanation for this, but it could be due to current loss in the bent transmission lines in the experiment, leading to a lower driving \(\mathbf{J}\times \mathbf{B}\) force on the coronal plasma.
For the purposes of this paper, we are interested only in the morphology of the layer predicted by resistive MHD, which is qualitatively the same in simulations run at lower currents which better reproduce the measured density.
As we will show below, it is likely that correct handling of the two-fluid terms in extended MHD is necessary to explain the exact morphology and density in these experiments.

\subsection{Reduced density in the reconnection layer}

Although these simulations do not reproduce the morphology of the layer, we can use them to assess the impact of the guide field on the layer density.
Detailed analysis of the pressure balance in these simulations (not presented) shows that the pile-up of the guide field contributes to the hollow layer structure seen in Fig. \ref{fig:ta_sims}a-c, as the increased magnetic pressure pushes plasma out from the centre of the layer.
Of course, because these simulations do not fully match our experiments, we cannot conclude that the structure of the guide field inside the layer is the same between our numerical and experimental observations, but the simulations suggest that guide field pile-up is an important factor in reducing the layer electron density, and one which we can attempt to measure in future experiments using end-on Faraday rotation imaging.\cite{Swadling2014a}

The reduction in electron density due to the line integration through the twisted reconnection layer is harder to evaluate from simulations, as we cannot turn off the reduction due to the guide field whilst keeping the twisted layer from the tilted array geometry.
However, a toy model described in Appendix \ref{sec:nel_twisted_layer} allows us to estimate the approximate reduction expected for the $b = 1$ case compared to $b=0$.
This model converts the line integration in $z$ to a convolution in $x$ of the layer density profile with a rectangular function, and finds a reduction in line-integrated density by a factor of around three.
Crude as this model is, it gives a rough estimate of how much we expect \neL to decrease simply due to geometric effects of the twisted layer on the line-integrated electron density maps.
Together with the guide field pile-up, we can qualitatively explain the reduced \neL in the layer seen in Fig. \ref{fig:ta_results}g.

\subsection{Morphology of the reconnection layer}

\begin{figure*}
    \includegraphics[width=1\linewidth]{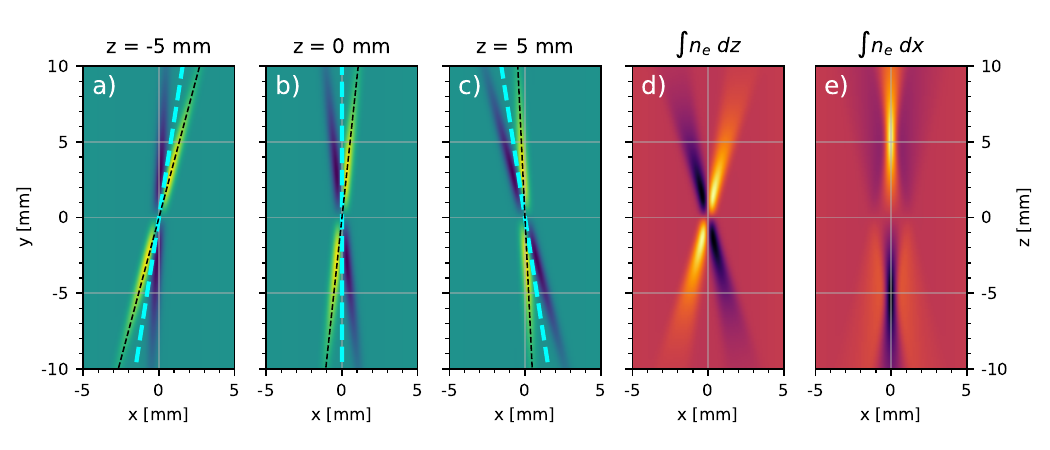}
    \caption{\label{fig:ta_2F_model}
    Results from our toy model combining the quadrupolar density structure seen in guide field ($b=1$) Hall reconnection, with our model for the geometric layer angle. 
    Panels (a--c) show slices in $n_e$ in the end-on $(x,y)$ plane, showing the quadrupolar density structure combined with the geometric model (eqn. \ref{eqn:layer_angle}) at (-5, 0, 5) mm, respectively. 
    The cyan dashed line indicates the geometric layer angle predicted by our analytical model, and the black dashed line indicates the predicted separatrix angle with respect to the geometric layer angle.
    Panel d) and e) show line-integrated views of the reconnection layer, in the end-on $(x, y)$ and side-on $(x, z)$ lines-of-sight, respectively. Both line-integrated structures show structures not seen in the GORGON 3D MHD simulations (Fig. \ref{fig:ta_sims}), but matching---qualitatively---those seen in the interferometry (Fig. \ref{fig:ta_results2}). 
    }
\end{figure*}

Unlike the experiments on MRX discussed in the introduction, we do not measure the out-of-plane magnetic fields in this experiment, and so we have no direct evidence for the quadrupolar magnetic field structure which is characteristic of Hall reconnection.
However, we can calculate the ion skin depth $d_i = c/\omega_{pi} \approx \SI{0.5}{\mm}$ for the ion densities $n_i = n_e/\bar{Z}$ inferred from the line-integrated electron densities shown in Fig. \ref{fig:ta_results}g, h and i.
The shaded region in Figs. \ref{fig:ta_results}g, h and i corresponds to $2d_i \gtrsim 2\Delta$, the reconnection layer width, which implies that the Hall term should be significant in our experiments.
As we also have a guide field, we therefore expect to see evidence for a quadrupolar density structure, as predicted by theory\cite{Kleva1995}, simulations,\cite{Huba2004, Yang2006} and seen in MRX experiments.\cite{Fox2017}

We consider the effects of the quadrupolar density structure on our line-integrated electron density maps using a toy model shown in Fig. \ref{fig:ta_2F_model}.
Here we consider a quadrupolar distribution with increased density along one separatrix (dashed black line) and decreased density along the other separatrix shown in Fig. \ref{fig:ta_2F_model}b.
This quadrupolar distribution is motivated especially by the simulation shown in Fig. 3 of Yang \textit{et al.}\cite{Yang2008}, which has a guide field ratio $b = 1$ relevant to our experiment.
We estimate the separatrix angle from the Sweet-Parker model, $\delta / L = S_L^{-1/2}$, assuming $S_L = 100$, based on previous experiments\cite{Hare2017}.
We build up a 3D model of the reconnection layer by rotating this quadrupolar distribution according to our geometric model (eqn. \ref{eqn:layer_angle}, blue dashed line in Fig. \ref{fig:ta_2F_model}), and we also show the density distributions at $z = \pm \SI{5}{\mm}$ in Fig. \ref{fig:ta_2F_model}a and c, respectively.

From this 3D model, we calculate the line-integrated electron density maps in the end-on and side-on configurations.
For the end-on model (Fig. \ref{fig:ta_2F_model}d), we see a quadrupolar pattern with an angle between the lower and higher density regions which is larger than the angle between the two separatrices in the $(x,y)$ slice of the density distribution (Fig. \ref{fig:ta_2F_model}b).
This is because the line integration through the twisted layer means that regions of higher density in one plane are cancelled out by corresponding regions of lower density in another plane.
The only remaining features come from the low-density separatrix at the lowest plane ($z = -\SI{5}{\mm}$, Fig. \ref{fig:ta_2F_model}a) and the high-density separatrix at the highest plane ($z = +\SI{5}{\mm}$, Fig. \ref{fig:ta_2F_model}b).
Qualitatively this quadrupolar density pattern reproduces that seen in our experimental data in Fig. \ref{fig:ta_results2}.

For the side-on model (Fig. \ref{fig:ta_2F_model}e), we see a more complex structure, which is left/right symmetric and up/down antisymmetric.
In the top half of the map, there is a dense region at $x = \SI{0}{\mm}$ flanked by two regions of depleted density.
In the bottom half of the map, there is a less dense region at the centre with higher-density regions on either side.
This is because in the bottom half of the map the low-density separatrix points predominantly along $x = \SI{0}{\mm}$ (see Fig. \ref{fig:ta_2F_model}a), and in the top half, the high-density separatrix points along $x = \SI{0}{\mm}$ (see Fig. \ref{fig:ta_2F_model}c).
Hence, the quadrupolar density structure also produces a unique signature in the side-on electron density map, which qualitatively reproduces the results in Fig. \ref{fig:ta_results2}.

Invoking two-fluid effects, which have been shown to combine with guide fields to produce a distinctive quadrupolar in-plane density structures, provides a convincing explanation for the structures seen in our end-on and side-on \neL maps.

\section{Conclusions and future work}

We have demonstrated a new pulsed-power-driven magnetic reconnection platform which directly embeds a guide field in the plasma flows using two tilted exploding wire arrays.
This resulted in a three-dimensional twisted layer structure, which significantly complicates the interpretation of line-integrated diagnostics.
By considering simple geometric models of the layer, backed by three-dimensional MHD simulations, we showed that the quadrupolar density structures seen in line-integrated electron density maps with a non-zero guide field could not be explained by resistive MHD.
We were able to qualitatively explain these structures by comparing our results to two-dimensional simulations from the literature, which include two-fluid effects (specifically the Hall term) and a guide field.

Our results extend previous work in two important ways.
Firstly, we observed the structure of the reconnection layer over larger length scales ($\Delta y = \SI{20}{\mm} \approx 40 d_i$) than accessible in previous experiments ($ 4\rho_s$ on MRX\cite{Fox2017}, $6d_i$ on CS-3D\cite{Frank2006}).
This enables us to study the global morphology introduced by the interaction of a guide field and the Hall field, which is especially relevant for understanding in situ measurements of reconnection layers made by spacecraft which cannot measure the global properties of the layer.\cite{Eriksson2016}

Secondly, previous work was carried out in the magnetically dominated regime ($\beta_{th}\equiv p_{th}/p_B\ll1$) in which the thermal and ram pressures are negligible.
In our experiments, $p_{th}\approx p_B\approx p_{ram}$ ($\beta_{th} \sim \beta_{kin}\sim 1$), which means that none of the pressure components can be neglected in the analysis of the layer dynamics. 
This regime is astrophysically relevant, as the long timescales for astrophysical plasmas lead to an approximate equipartition between the energy density components.
As such, our experiments extend the regimes of reconnection in which the interaction between two-fluid effects and the guide field can be studied.

Our results, however, are confined to measurements only of the electron density. 
In order to understand the full impact of the guide field on two-fluid reconnection, we need to measure all of the relevant quantities, including the magnetic field, the plasma flow velocity, and the electron and ion temperatures.
These measurements can be made using Faraday rotation imaging and Thomson scattering \cite{Swadling2014a, Hare2017}, but the three-dimensional structure of the reconnection layer will complicate the interpretation of these results.
As such, in future work, we aim to design new platforms for generating plasma flows with an embedded guide field which will produce a reconnection layer with more straightforward symmetries.

\section{Acknowledgements}
We gratefully acknowledge the assistance of Prof. L. Alexander Betts with the symmetry and analytical rocket model arguments. 
This work was funded by NSF and NNSA under grant no. PHY2108050, and also supported by the NNSA Stewardship Science Academic Programs under DOE Cooperative Agreement DE-NA0004148.
TWOV acknowledges support from the MIT MathWorks fellowship. 
We would also like to thank Chris Haynes and the rest of the Morningside Academy for Design (MAD) for their help with the fabrication of the tilted array load hardware, along with the engineering team at the MIT Plasma Science and Fusion Center (PSFC) for their work machining the load hardware.
We are grateful to Prof. Jeremy Chittenden for providing us with the GORGON code, and for helping us set up these simulations.

\section{Declaration of Conflicts of Interest}

The authors have no conflicts of interest to disclose.

\section{Data Availability}

The data that support the findings of this study are available from the corresponding author upon reasonable request.

\appendix
\section*{Appendices}
\section{Interferometry Zeroing Method}\label{sec:zeroing}
Laser interferometry provides a relative measurement of the line-integrated electron density, which can be an absolute measurement if there is a region with no plasma, and hence no fringe distortion.
This zero fringe shift region provides a reference point from which we can measure the absolute line-integrated electron density.
Due to the limited field of view, not all of the interferograms have such a region.
However, there is a reliable region for the side-on, $b = 0$ interferogram.
Using this, and the assumption that the tilted arrays will ablate the same amount of plasma regardless of tilt angle, we have developed a method for estimating the zero of the line-integrated electron density scale for the other \neL maps.
This ensures that the \neL maps are approximately comparable, especially for visual comparisons with shared colourmaps.
The procedure is as follows:
\begin{enumerate}
    \item Taking the side-on $\langle n_e L_y \rangle$ map of the $b=0$ shot (m2986), we integrate $\int \left( \langle n_e L_y \rangle \right) dz$. Taking the end-on $\langle n_e L_z \rangle$ map of that same shot, we integrate $\int \left( \langle n_e L_z \rangle \right) dy$. The resulting profiles are both $\int n_e(x) dydz$.
    \item In the ideal case where our diagnostic has captured the entire extent of the plasma, these profiles should match. However, our limited field-of-view means that there are regions of plasma which are not included in each line-of-sight. 
    To correct for this, we add a positive offset to the \neL map with the lowest value of $\int n_e(x) dydz$ near the peak centred between the two arrays. Adding a positive offset avoids any negative values of \neL in either map. 
    We now have a set of line-integrated electron density maps for the untilted hardware which have consistent densities within the reconnection layer.
    \item For the $b=$ 0.4 and 1 \neL maps, there is no zero fringe shift region in the side-on or end-on maps.
    However, we can zero these by requiring consistency with the $b = 0$ case (m2986).
    As the ablation rate from the wires should be independent of tilt angle, we compare the density in the ablation flows, accounting for the tilted geometry and differing peak currents between shots.
    We compare lineouts of the side-on \neL map along $z=0$ for the $b=0$ and $b\neq0$ shots. 
    To account for the different shot currents, we divide by the square of the ratio of the peak shot current to the reference ($b=0$) shot current:
    $$\left( \frac{I_{peak,b\neq0}}{I_{peak,b=0}} \right)^2.$$
    This current scaling factor is derived from the ablation rate predicted by the rocket model.\cite{Lebedev2001}
    We also account for the geometric effect on the line-integrated density introduced by tilting the arrays. 
    For a tilted array, the integration path is longer than for an upright array, and so we divide by \(\sqrt{1 + \tan^2(\theta)}\),
    where $\theta$ is the tilt angle of the array. 
    With both scalings applied, the line integrated density at $|x|>\SI{3}{\mm}$ is matched by applying an offset to $b\neq0$ \neL map. 
    We apply an additional offset to ensure that the map contains no negative densities.
    \item Now that the side-on $b\neq0$ map is consistent with the side-on $b=0$ map, we can zero the $b\neq0$ end-on map with the corresponding $b\neq0$ side-on map, by following steps 1--2 above.
\end{enumerate}

% \textcolor{red}{zeroing method (draft, in-progress)}
% \begin{itemize}
%     \item (Taking m2986, zero deg.) Integrate side-on and end-on ``down'' so that we have something we can compare between the two: $\int n_e \ dy \ dz$. This ends up as a sort of ``lineout'' profile, as a function of $x$. In ideal limit, should be same.
%     \item Match the layer peaks between the side-on/end-on.
%     \item Since side-on is lower than end-on, we add on the shift to the side-on (to prevent there being negative values by subtracting from the end-on). Shift by a constant.
%     \item Now, taking a new shot (e.g. 22.5 deg. 2996)... we will compare it with the 0 deg. side-on. Plot the side-on for both, side-by-side. Take lineouts of each at $z=0$, BUT for the tilted case, we divide the lineout by (geometric scaling multiplied by current scaling), and add a shift. We want to match the inflow regions (i.e. $\abs{x} > 3$). Shift specified manually ($\sim 1.3e17$). Geometric scaling: $\sqrt{1 + \tan^2(22.5 \si{deg})}$. Current scaling: $$(I_{peak,2996} / I_{peak,2986})^2$$. Current scaling derived from the rocket model. \textcolor{red}{Where did the geometric scaling come from?}.
%     \item Now that the side-on is matched (geometrically/current) to the 0 deg. case... we shift the end-on UP to match the side-on layer peak (like was done with the 0 deg. case).
%     \item This comparison is repeated for each angle (22.5, 45), comparing always back to the 0 deg. case, making sure to apply the relevant scaling.
% \end{itemize}
% For b!=0, don't have good zero fringe shift.
% This should be appendix.

\section{A Symmetry Argument for the \texorpdfstring{$\langle n_e L \rangle$}{neL} Maps}\label{sec:symmetry}

Our load hardware (Fig. \ref{fig:ta_setup}) has two-fold (\SI{180}{\deg}) rotational symmetry around the principal axes $x$, $y$, and $z$.
That is, the lines describing the axes (central cathode posts) of the two arrays can be described by a function $f(x,y,z) = f(-x, -y, z) = f(x, -y, -z) = f(-x, y, -z)$. 
This can be easily verified by considering the coordinates of the top and the bottom of each array: for the $b = 1$ case shown in Fig \ref{fig:ta_setup} these are $[1,1,1]$, $[-1,-1,1]$, $[-1,1,-1]$ and $[1,-1,-1]$, with appropriate scaling.

In the absence of any symmetry breaking effects such as the Hall term, we would expect that the electron density $n_e(x,y,z)$ should have the same symmetry as the load hardware which produces it, and so integrations of the electron density along $x$, $y$ or $z$ will yield line-integrated electron density maps with symmetry.
For example 
\begin{align}
    \int n_e(x,y,z) dz =& \int n_e(x,-y,z) dz \nonumber\\
    =& \int n_e(-x,y,-z) dz \nonumber\\
    =& \int n_e(-x,-y,- z) dz,
\end{align} 

and hence
\begin{align}
    \langle n_e L_z\rangle (x,y)=& \langle n_e L_z\rangle (x,-y) \nonumber\\
    =&\langle n_e L_z\rangle (-x,y)\nonumber\\
    =&\langle n_e L_z\rangle (-x,-y).
\end{align}
Therefore the two-dimensional line-integrated density maps exhibit mirror symmetries around the remaining two axes---for $\langle n_e L_z\rangle$ (the end-on view) there are mirror symmetries around $x = 0$ and $y = 0$, and for $\langle n_e L_y\rangle$ (the side-on view) around $x = 0$ and $z = 0$.

Rather than these strict symmetries expected from our load hardware, our experimental results show quadrupolar structures and up/down antisymmetry, which implies that in these experiments there are plasma physics effects which can break the symmetry, such as the Hall term.

\section{A Geometric Model For A Twisted Reconnection Layer} \label{app:geometric_model}
Using the well-known rocket model for ablation from a wire array \cite{Lebedev2001}, we derive a model for the geometric shape of the reconnection layer.
Figure \ref{fig:ta_model} illustrates the setup of this model.
First, we need to find an expression for the radial distance to a point $(x, y, z)$ from the tilted central axis of the array.
For array 2, centred around $(x_0, 0, 0)$, and tilted by $\theta$ from the vertical ($z$) axis, the radial distance $r_2$ is:
\begin{align}
    r_2^2 &= (x-x_0)^2 + (y-y_0)^2 + (z-z_0)^2, \\
    \text{where } z_0 &= \ell_2 \cos\theta,\quad  y_0 = \ell_2 \sin\theta
\end{align}
and $\ell_2$ is the distance from $(x_0, 0, 0)$ to $(x_0, y_0, z_0)$, which is the closest point on the array axis to $(x, y, z)$
\begin{equation}
    \ell_2 = y \sin\theta + z \cos\theta.
\end{equation}
Similarly, for the first array, centred around $(-x_0, 0, 0)$ and tilted by $-\theta$ from the $z$ axis,
\begin{align}
    r_1^2 &= (x+x_0)^2 + (y+\ell_1\sin\theta)^2 + (z-\ell_1\cos\theta)^2 \\
    \ell_1 &= z \cos\theta - y \sin\theta.
\end{align}
We can relate $r_1$ to $r_2$ using the rocket model\cite{Lebedev2001}, which describes the mass density around a single exploding wire array of radius $R_0$ driven by a current $I(t)$, as a function of ablation velocity $V_{abl}$ (assumed constant), $r$, and time $t$:
\begin{align}
    \rho(r,t)\propto \frac{1}{rR_0V_{abl}} \left[ I^2\left(t-\frac{R_0-r}{V_{abl}}\right) \right].
\end{align}
We hypothesise that the reconnection layer will form where the mass density in the plasma flows from each array is equal: \(\rho(r_1) = \rho(r_2)\).
This condition is satisfied when $r_1 = r_2$, and at this location, the additional time-of-flight effects encoded in $t-(R_0-r)/V_{abl}$ are equal, and so the model is time-invariant.

We can now equate our separate equations for $r_1^2$ and $r_2^2$, and simplify:
\begin{equation}
    2x_0 \cdot x = -\sin(2\theta) \cdot y \cdot z.
\end{equation}
Recall, as we tilt the exploding wire arrays by an angle, $\theta$, we increase the guide field ratio $b = \frac{B_{g}}{B_{rec}} = \tan\theta$.
Therefore, we can rewrite this in terms of $b$, and rearrange for $y$:
\begin{equation}
    y = -\frac{x_0 (1 + b^2)}{b} \cdot \frac{1}{z} \cdot x.
\end{equation}
This equation describes a three-dimensional structure, which is a doubly-ruled surface known as a hyperbolic paraboloid.

\section{A Model for the Reduction in Line-Integrated Electron Density in a Twisted Layer}\label{sec:nel_twisted_layer}

In this model, we consider a peaked density profile in $x$ at some position $y = y_0$ in the plane $z = 0$, for example 
\begin{align}
    n_e(x, y = y_0, z = 0) = n_{e0} \exp(-x^2/\delta^2),
\end{align}
which has a density peak of $n_{e0}$ and a width of $\delta$.
In planes other than $z = 0$, this profile will be centred on 
\begin{align}\label{eqn:xtoz}
    x' = -\frac{y_0 z}{x_0} \frac{b}{1+b^2},
\end{align} 
due to the geometric shape of the layer in three dimensions (see eqn. \ref{eqn:layer_angle}), and hence \begin{align}
    n_e(x) = n_{e0} \exp(-(x-x')^2/\delta^2).
\end{align}
We can calculate the line-integrated density \begin{align}
    \langle n_eL_z \rangle(x=0, y = y_0) =  \int_{-z_0}^{+z_0} n_e(x = 0, y = y_0, z) dz,
\end{align}
and compare between the $b = 0$ and $b = 1$ cases to estimate the contribution from these geometric effects.
For $b = 0$ the result is simply \(\langle n_eL_z \rangle(x=0, y = y_0) = 2 z_0 n_{e0}\), as expected.
For $b=1$ the integral \(\int_{-z_0}^{+z_0} n_e(x-x') dz\) can be written as a convolution by changing variables from $z$ to $x'$ using eqn \ref{eqn:xtoz}: 
\begin{equation}
    \int n_e(x-x') \Pi(x'/2 x_M) \cdot (-2x_0/y_0) dx'
\end{equation} where $\Pi(x'/2x_M)$ is a rectangular function with width $2 x_M = -y_0 z_0/x_0$, which limits the domain of integration from $[-\infty, \infty]$ to $[-x_M, x_M]$.
The result of this convolution is a standard result, 
\begin{align}
    \langle n_eL_z \rangle(x, y = y_0) = & 2 \sqrt{\pi} n_{e0} \delta x_0/y_0 \\
    &\times \{\erf[(x+x_M)/\delta]-\erf[(x-x_M)/\delta]\}\nonumber 
\end{align}
For $x_M\gg\delta$, which is typical for our experiments, the difference between the two error functions is approximately $2 \Pi(x/2x_M)$, and so this toy model predicts that the peak line-integrated density:
\begin{equation}
    \langle n_eL_z \rangle(x = 0, y = y_0) = 2 \sqrt{\pi} \delta (x_0/y_0) n_{e0}.\label{eqn:nel_twisted_layer}
\end{equation}
In our experiment, the geometry sets $x_0 = \SI{11}{\mm}$, and we observe \(\delta \approx \SI{0.5}{\mm}\) (Fig. \ref{fig:ta_results}g).
As reference point we consider $y_0 = \SI{5}{mm}$, and so for $b = 1$, the predicted line-integrated density here is $\approx4 n_{e0}$.
In comparison, for the $b = 0$ case we calculate $12 n_{e0}$, which implies that the rotation of the layer reduces the measured line-integrated electron density by a factor of three.
%see https://math.stackexchange.com/questions/272001/double-tophat-convolved-with-a-gaussian for the convolution

\bibliography{library,references}

\providecommand{\noopsort}[1]{}\providecommand{\singleletter}[1]{#1}%
\begin{thebibliography}{35}%
\makeatletter
\providecommand \@ifxundefined [1]{%
 \@ifx{#1\undefined}
}%
\providecommand \@ifnum [1]{%
 \ifnum #1\expandafter \@firstoftwo
 \else \expandafter \@secondoftwo
 \fi
}%
\providecommand \@ifx [1]{%
 \ifx #1\expandafter \@firstoftwo
 \else \expandafter \@secondoftwo
 \fi
}%
\providecommand \natexlab [1]{#1}%
\providecommand \enquote  [1]{``#1''}%
\providecommand \bibnamefont  [1]{#1}%
\providecommand \bibfnamefont [1]{#1}%
\providecommand \citenamefont [1]{#1}%
\providecommand \href@noop [0]{\@secondoftwo}%
\providecommand \href [0]{\begingroup \@sanitize@url \@href}%
\providecommand \@href[1]{\@@startlink{#1}\@@href}%
\providecommand \@@href[1]{\endgroup#1\@@endlink}%
\providecommand \@sanitize@url [0]{\catcode `\\12\catcode `\$12\catcode `\&12\catcode `\#12\catcode `\^12\catcode `\_12\catcode `\%12\relax}%
\providecommand \@@startlink[1]{}%
\providecommand \@@endlink[0]{}%
\providecommand \url  [0]{\begingroup\@sanitize@url \@url }%
\providecommand \@url [1]{\endgroup\@href {#1}{\urlprefix }}%
\providecommand \urlprefix  [0]{URL }%
\providecommand \Eprint [0]{\href }%
\providecommand \doibase [0]{http://dx.doi.org/}%
\providecommand \selectlanguage [0]{\@gobble}%
\providecommand \bibinfo  [0]{\@secondoftwo}%
\providecommand \bibfield  [0]{\@secondoftwo}%
\providecommand \translation [1]{[#1]}%
\providecommand \BibitemOpen [0]{}%
\providecommand \bibitemStop [0]{}%
\providecommand \bibitemNoStop [0]{.\EOS\space}%
\providecommand \EOS [0]{\spacefactor3000\relax}%
\providecommand \BibitemShut  [1]{\csname bibitem#1\endcsname}%
\let\auto@bib@innerbib\@empty
%</preamble>
\bibitem [{\citenamefont {Zweibel}\ and\ \citenamefont {Yamada}(2009)}]{Zweibel2009}%
  \BibitemOpen
  \bibfield  {author} {\bibinfo {author} {\bibfnamefont {E.}~\bibnamefont {Zweibel}}\ and\ \bibinfo {author} {\bibfnamefont {M.}~\bibnamefont {Yamada}},\ }\href {\doibase 10.1146/annurev-astro-082708-101726} {\bibfield  {journal} {\bibinfo  {journal} {Annual Review of Astronomy and Astrophysics}\ }\textbf {\bibinfo {volume} {47}},\ \bibinfo {pages} {291} (\bibinfo {year} {2009})}\BibitemShut {NoStop}%
\bibitem [{\citenamefont {Zweibel}\ and\ \citenamefont {Yamada}(2016)}]{Zweibel2016}%
  \BibitemOpen
  \bibfield  {author} {\bibinfo {author} {\bibfnamefont {E.~G.}\ \bibnamefont {Zweibel}}\ and\ \bibinfo {author} {\bibfnamefont {M.}~\bibnamefont {Yamada}},\ }\href {\doibase 10.1098/rspa.2016.0479} {\bibfield  {journal} {\bibinfo  {journal} {Proceedings of the Royal Society A: Mathematical, Physical and Engineering Science}\ }\textbf {\bibinfo {volume} {472}},\ \bibinfo {pages} {20160479} (\bibinfo {year} {2016})}\BibitemShut {NoStop}%
\bibitem [{\citenamefont {Huba}(2004)}]{Huba2004}%
  \BibitemOpen
  \bibfield  {author} {\bibinfo {author} {\bibfnamefont {J.~D.}\ \bibnamefont {Huba}},\ }\href {\doibase 10.1063/1.1834592} {\bibfield  {journal} {\bibinfo  {journal} {Physics of Plasmas}\ }\textbf {\bibinfo {volume} {12}},\ \bibinfo {pages} {012322} (\bibinfo {year} {2004})}\BibitemShut {NoStop}%
\bibitem [{\citenamefont {Yang}, \citenamefont {Jin},\ and\ \citenamefont {Zhou}(2006)}]{Yang2006}%
  \BibitemOpen
  \bibfield  {author} {\bibinfo {author} {\bibfnamefont {H.}~\bibnamefont {Yang}}, \bibinfo {author} {\bibfnamefont {S.}~\bibnamefont {Jin}}, \ and\ \bibinfo {author} {\bibfnamefont {G.}~\bibnamefont {Zhou}},\ }\href {\doibase 10.1029/2005JA011536} {\bibfield  {journal} {\bibinfo  {journal} {Journal of Geophysical Research (Space Physics)}\ }\textbf {\bibinfo {volume} {111}},\ \bibinfo {pages} {11223} (\bibinfo {year} {2006})}\BibitemShut {NoStop}%
\bibitem [{\citenamefont {Cassak}\ \emph {et~al.}(2017)\citenamefont {Cassak}, \citenamefont {Genestreti}, \citenamefont {Burch}, \citenamefont {Phan}, \citenamefont {Shay}, \citenamefont {Swisdak}, \citenamefont {Drake}, \citenamefont {Price}, \citenamefont {Eriksson}, \citenamefont {Ergun}, \citenamefont {Anderson}, \citenamefont {Merkin},\ and\ \citenamefont {Komar}}]{Cassak2017}%
  \BibitemOpen
  \bibfield  {author} {\bibinfo {author} {\bibfnamefont {P.~A.}\ \bibnamefont {Cassak}}, \bibinfo {author} {\bibfnamefont {K.~J.}\ \bibnamefont {Genestreti}}, \bibinfo {author} {\bibfnamefont {J.~L.}\ \bibnamefont {Burch}}, \bibinfo {author} {\bibfnamefont {T.-D.}\ \bibnamefont {Phan}}, \bibinfo {author} {\bibfnamefont {M.~A.}\ \bibnamefont {Shay}}, \bibinfo {author} {\bibfnamefont {M.}~\bibnamefont {Swisdak}}, \bibinfo {author} {\bibfnamefont {J.~F.}\ \bibnamefont {Drake}}, \bibinfo {author} {\bibfnamefont {L.}~\bibnamefont {Price}}, \bibinfo {author} {\bibfnamefont {S.}~\bibnamefont {Eriksson}}, \bibinfo {author} {\bibfnamefont {R.~E.}\ \bibnamefont {Ergun}}, \bibinfo {author} {\bibfnamefont {B.~J.}\ \bibnamefont {Anderson}}, \bibinfo {author} {\bibfnamefont {V.~G.}\ \bibnamefont {Merkin}}, \ and\ \bibinfo {author} {\bibfnamefont {C.~M.}\ \bibnamefont {Komar}},\ }\href {\doibase 10.1002/2017JA024555} {\bibfield  {journal} {\bibinfo  {journal} {Journal of Geophysical Research: Space Physics}\ }\textbf
  {\bibinfo {volume} {122}},\ \bibinfo {pages} {11,523} (\bibinfo {year} {2017})}\BibitemShut {NoStop}%
\bibitem [{\citenamefont {Egedal}, \citenamefont {Daughton},\ and\ \citenamefont {Le}(2012)}]{Egedal2012}%
  \BibitemOpen
  \bibfield  {author} {\bibinfo {author} {\bibfnamefont {J.}~\bibnamefont {Egedal}}, \bibinfo {author} {\bibfnamefont {W.}~\bibnamefont {Daughton}}, \ and\ \bibinfo {author} {\bibfnamefont {A.}~\bibnamefont {Le}},\ }\href {\doibase 10.1038/nphys2249} {\bibfield  {journal} {\bibinfo  {journal} {Nature Physics}\ }\textbf {\bibinfo {volume} {8}},\ \bibinfo {pages} {321} (\bibinfo {year} {2012})}\BibitemShut {NoStop}%
\bibitem [{\citenamefont {Phan}\ \emph {et~al.}(2010)\citenamefont {Phan}, \citenamefont {Gosling}, \citenamefont {Paschmann}, \citenamefont {Pasma}, \citenamefont {Drake}, \citenamefont {{\O}ieroset}, \citenamefont {Larson}, \citenamefont {Lin},\ and\ \citenamefont {Davis}}]{Phan2010}%
  \BibitemOpen
  \bibfield  {author} {\bibinfo {author} {\bibfnamefont {T.~D.}\ \bibnamefont {Phan}}, \bibinfo {author} {\bibfnamefont {J.~T.}\ \bibnamefont {Gosling}}, \bibinfo {author} {\bibfnamefont {G.}~\bibnamefont {Paschmann}}, \bibinfo {author} {\bibfnamefont {C.}~\bibnamefont {Pasma}}, \bibinfo {author} {\bibfnamefont {J.~F.}\ \bibnamefont {Drake}}, \bibinfo {author} {\bibfnamefont {M.}~\bibnamefont {{\O}ieroset}}, \bibinfo {author} {\bibfnamefont {D.}~\bibnamefont {Larson}}, \bibinfo {author} {\bibfnamefont {R.~P.}\ \bibnamefont {Lin}}, \ and\ \bibinfo {author} {\bibfnamefont {M.~S.}\ \bibnamefont {Davis}},\ }\href {\doibase 10.1088/2041-8205/719/2/L199} {\bibfield  {journal} {\bibinfo  {journal} {The Astrophysical Journal Letters}\ }\textbf {\bibinfo {volume} {719}},\ \bibinfo {pages} {L199} (\bibinfo {year} {2010})}\BibitemShut {NoStop}%
\bibitem [{\citenamefont {Eastwood}\ \emph {et~al.}(2010)\citenamefont {Eastwood}, \citenamefont {Shay}, \citenamefont {Phan},\ and\ \citenamefont {{\O}ieroset}}]{Eastwood2010}%
  \BibitemOpen
  \bibfield  {author} {\bibinfo {author} {\bibfnamefont {J.~P.}\ \bibnamefont {Eastwood}}, \bibinfo {author} {\bibfnamefont {M.~A.}\ \bibnamefont {Shay}}, \bibinfo {author} {\bibfnamefont {T.~D.}\ \bibnamefont {Phan}}, \ and\ \bibinfo {author} {\bibfnamefont {M.}~\bibnamefont {{\O}ieroset}},\ }\href {\doibase 10.1103/PhysRevLett.104.205001} {\bibfield  {journal} {\bibinfo  {journal} {Physical Review Letters}\ }\textbf {\bibinfo {volume} {104}},\ \bibinfo {pages} {205001} (\bibinfo {year} {2010})}\BibitemShut {NoStop}%
\bibitem [{\citenamefont {Uzdensky}\ and\ \citenamefont {Kulsrud}(2006)}]{Uzdensky2006a}%
  \BibitemOpen
  \bibfield  {author} {\bibinfo {author} {\bibfnamefont {D.~A.}\ \bibnamefont {Uzdensky}}\ and\ \bibinfo {author} {\bibfnamefont {R.~M.}\ \bibnamefont {Kulsrud}},\ }\href {\doibase 10.1063/1.2209627} {\bibfield  {journal} {\bibinfo  {journal} {Physics of Plasmas}\ }\textbf {\bibinfo {volume} {13}},\ \bibinfo {pages} {062305} (\bibinfo {year} {2006})}\BibitemShut {NoStop}%
\bibitem [{\citenamefont {Drake}\ and\ \citenamefont {Burkhart}(1992)}]{Drake1992a}%
  \BibitemOpen
  \bibfield  {author} {\bibinfo {author} {\bibfnamefont {J.~F.}\ \bibnamefont {Drake}}\ and\ \bibinfo {author} {\bibfnamefont {G.~R.}\ \bibnamefont {Burkhart}},\ }\href {\doibase 10.1029/92GL01034} {\bibfield  {journal} {\bibinfo  {journal} {Geophysical Research Letters}\ }\textbf {\bibinfo {volume} {19}},\ \bibinfo {pages} {1077} (\bibinfo {year} {1992})}\BibitemShut {NoStop}%
\bibitem [{\citenamefont {Ren}\ \emph {et~al.}(2005)\citenamefont {Ren}, \citenamefont {Yamada}, \citenamefont {Gerhardt}, \citenamefont {Ji}, \citenamefont {Kulsrud},\ and\ \citenamefont {Kuritsyn}}]{Ren2005a}%
  \BibitemOpen
  \bibfield  {author} {\bibinfo {author} {\bibfnamefont {Y.}~\bibnamefont {Ren}}, \bibinfo {author} {\bibfnamefont {M.}~\bibnamefont {Yamada}}, \bibinfo {author} {\bibfnamefont {S.}~\bibnamefont {Gerhardt}}, \bibinfo {author} {\bibfnamefont {H.}~\bibnamefont {Ji}}, \bibinfo {author} {\bibfnamefont {R.}~\bibnamefont {Kulsrud}}, \ and\ \bibinfo {author} {\bibfnamefont {A.}~\bibnamefont {Kuritsyn}},\ }\href {\doibase 10.1103/PhysRevLett.95.055003} {\bibfield  {journal} {\bibinfo  {journal} {Physical Review Letters}\ }\textbf {\bibinfo {volume} {95}},\ \bibinfo {pages} {1} (\bibinfo {year} {2005})}\BibitemShut {NoStop}%
\bibitem [{\citenamefont {Yamada}\ \emph {et~al.}(2006)\citenamefont {Yamada}, \citenamefont {Ren}, \citenamefont {Ji}, \citenamefont {Breslau}, \citenamefont {Gerhardt}, \citenamefont {Kulsrud},\ and\ \citenamefont {Kuritsyn}}]{Yamada2006}%
  \BibitemOpen
  \bibfield  {author} {\bibinfo {author} {\bibfnamefont {M.}~\bibnamefont {Yamada}}, \bibinfo {author} {\bibfnamefont {Y.}~\bibnamefont {Ren}}, \bibinfo {author} {\bibfnamefont {H.}~\bibnamefont {Ji}}, \bibinfo {author} {\bibfnamefont {J.}~\bibnamefont {Breslau}}, \bibinfo {author} {\bibfnamefont {S.}~\bibnamefont {Gerhardt}}, \bibinfo {author} {\bibfnamefont {R.}~\bibnamefont {Kulsrud}}, \ and\ \bibinfo {author} {\bibfnamefont {A.}~\bibnamefont {Kuritsyn}},\ }\href {\doibase 10.1063/1.2203950} {\bibfield  {journal} {\bibinfo  {journal} {Physics of Plasmas}\ }\textbf {\bibinfo {volume} {13}},\ \bibinfo {pages} {052119} (\bibinfo {year} {2006})}\BibitemShut {NoStop}%
\bibitem [{\citenamefont {Mozer}, \citenamefont {Bale},\ and\ \citenamefont {Phan}(2002)}]{Mozer2002}%
  \BibitemOpen
  \bibfield  {author} {\bibinfo {author} {\bibfnamefont {F.~S.}\ \bibnamefont {Mozer}}, \bibinfo {author} {\bibfnamefont {S.~D.}\ \bibnamefont {Bale}}, \ and\ \bibinfo {author} {\bibfnamefont {T.~D.}\ \bibnamefont {Phan}},\ }\href {\doibase 10.1103/PhysRevLett.89.015002} {\bibfield  {journal} {\bibinfo  {journal} {Physical Review Letters}\ }\textbf {\bibinfo {volume} {89}},\ \bibinfo {pages} {015002} (\bibinfo {year} {2002})}\BibitemShut {NoStop}%
\bibitem [{\citenamefont {Wygant}\ \emph {et~al.}(2005)\citenamefont {Wygant}, \citenamefont {Cattell}, \citenamefont {Lysak}, \citenamefont {Song}, \citenamefont {Dombeck}, \citenamefont {McFadden}, \citenamefont {Mozer}, \citenamefont {Carlson}, \citenamefont {Parks}, \citenamefont {Lucek}, \citenamefont {Balogh}, \citenamefont {Andre}, \citenamefont {Rerhe}, \citenamefont {Hesse},\ and\ \citenamefont {Mouikis}}]{Wygant2005}%
  \BibitemOpen
  \bibfield  {author} {\bibinfo {author} {\bibfnamefont {J.~R.}\ \bibnamefont {Wygant}}, \bibinfo {author} {\bibfnamefont {C.~A.}\ \bibnamefont {Cattell}}, \bibinfo {author} {\bibfnamefont {R.}~\bibnamefont {Lysak}}, \bibinfo {author} {\bibfnamefont {Y.}~\bibnamefont {Song}}, \bibinfo {author} {\bibfnamefont {J.}~\bibnamefont {Dombeck}}, \bibinfo {author} {\bibfnamefont {J.}~\bibnamefont {McFadden}}, \bibinfo {author} {\bibfnamefont {F.~S.}\ \bibnamefont {Mozer}}, \bibinfo {author} {\bibfnamefont {C.~W.}\ \bibnamefont {Carlson}}, \bibinfo {author} {\bibfnamefont {G.}~\bibnamefont {Parks}}, \bibinfo {author} {\bibfnamefont {E.~A.}\ \bibnamefont {Lucek}}, \bibinfo {author} {\bibfnamefont {A.}~\bibnamefont {Balogh}}, \bibinfo {author} {\bibfnamefont {M.}~\bibnamefont {Andre}}, \bibinfo {author} {\bibfnamefont {H.}~\bibnamefont {Rerhe}}, \bibinfo {author} {\bibfnamefont {M.}~\bibnamefont {Hesse}}, \ and\ \bibinfo {author} {\bibfnamefont {C.}~\bibnamefont {Mouikis}},\ }\href {\doibase 10.1029/2004JA010708}
  {\bibfield  {journal} {\bibinfo  {journal} {Journal of Geophysical Research: Space Physics}\ }\textbf {\bibinfo {volume} {110}},\ \bibinfo {pages} {1} (\bibinfo {year} {2005})}\BibitemShut {NoStop}%
\bibitem [{\citenamefont {Yang}, \citenamefont {Jin},\ and\ \citenamefont {Liu}(2008)}]{Yang2008}%
  \BibitemOpen
  \bibfield  {author} {\bibinfo {author} {\bibfnamefont {H.}~\bibnamefont {Yang}}, \bibinfo {author} {\bibfnamefont {S.}~\bibnamefont {Jin}}, \ and\ \bibinfo {author} {\bibfnamefont {C.}~\bibnamefont {Liu}},\ }\href {\doibase 10.1016/j.asr.2007.09.035} {\bibfield  {journal} {\bibinfo  {journal} {Advances in Space Research}\ }\textbf {\bibinfo {volume} {41}},\ \bibinfo {pages} {1649} (\bibinfo {year} {2008})}\BibitemShut {NoStop}%
\bibitem [{\citenamefont {Hare}\ \emph {et~al.}(2017{\natexlab{a}})\citenamefont {Hare}, \citenamefont {Lebedev}, \citenamefont {Suttle}, \citenamefont {Loureiro}, \citenamefont {Ciardi}, \citenamefont {Burdiak}, \citenamefont {Chittenden}, \citenamefont {Clayson}, \citenamefont {Eardley}, \citenamefont {Garcia}, \citenamefont {Halliday}, \citenamefont {Niasse}, \citenamefont {Robinson}, \citenamefont {Smith}, \citenamefont {Stuart}, \citenamefont {{Suzuki-Vidal}}, \citenamefont {Swadling}, \citenamefont {Ma},\ and\ \citenamefont {Wu}}]{Hare2017c}%
  \BibitemOpen
  \bibfield  {author} {\bibinfo {author} {\bibfnamefont {J.~D.}\ \bibnamefont {Hare}}, \bibinfo {author} {\bibfnamefont {S.~V.}\ \bibnamefont {Lebedev}}, \bibinfo {author} {\bibfnamefont {L.~G.}\ \bibnamefont {Suttle}}, \bibinfo {author} {\bibfnamefont {N.~F.}\ \bibnamefont {Loureiro}}, \bibinfo {author} {\bibfnamefont {A.}~\bibnamefont {Ciardi}}, \bibinfo {author} {\bibfnamefont {G.~C.}\ \bibnamefont {Burdiak}}, \bibinfo {author} {\bibfnamefont {J.~P.}\ \bibnamefont {Chittenden}}, \bibinfo {author} {\bibfnamefont {T.}~\bibnamefont {Clayson}}, \bibinfo {author} {\bibfnamefont {S.~J.}\ \bibnamefont {Eardley}}, \bibinfo {author} {\bibfnamefont {C.}~\bibnamefont {Garcia}}, \bibinfo {author} {\bibfnamefont {J.~W.~D.}\ \bibnamefont {Halliday}}, \bibinfo {author} {\bibfnamefont {N.}~\bibnamefont {Niasse}}, \bibinfo {author} {\bibfnamefont {T.}~\bibnamefont {Robinson}}, \bibinfo {author} {\bibfnamefont {R.~A.}\ \bibnamefont {Smith}}, \bibinfo {author} {\bibfnamefont {N.}~\bibnamefont {Stuart}}, \bibinfo {author}
  {\bibfnamefont {F.}~\bibnamefont {{Suzuki-Vidal}}}, \bibinfo {author} {\bibfnamefont {G.~F.}\ \bibnamefont {Swadling}}, \bibinfo {author} {\bibfnamefont {J.}~\bibnamefont {Ma}}, \ and\ \bibinfo {author} {\bibfnamefont {J.}~\bibnamefont {Wu}},\ }\href {\doibase 10.1063/1.4986012} {\bibfield  {journal} {\bibinfo  {journal} {Physics of Plasmas}\ }\textbf {\bibinfo {volume} {24}},\ \bibinfo {pages} {102703} (\bibinfo {year} {2017}{\natexlab{a}})}\BibitemShut {NoStop}%
\bibitem [{\citenamefont {Kleva}, \citenamefont {Drake},\ and\ \citenamefont {Waelbroeck}(1995)}]{Kleva1995}%
  \BibitemOpen
  \bibfield  {author} {\bibinfo {author} {\bibfnamefont {R.~G.}\ \bibnamefont {Kleva}}, \bibinfo {author} {\bibfnamefont {J.~F.}\ \bibnamefont {Drake}}, \ and\ \bibinfo {author} {\bibfnamefont {F.~L.}\ \bibnamefont {Waelbroeck}},\ }\href {\doibase 10.1063/1.871095} {\bibfield  {journal} {\bibinfo  {journal} {Physics of Plasmas}\ }\textbf {\bibinfo {volume} {2}},\ \bibinfo {pages} {23} (\bibinfo {year} {1995})}\BibitemShut {NoStop}%
\bibitem [{\citenamefont {Eriksson}\ \emph {et~al.}(2016)\citenamefont {Eriksson}, \citenamefont {Wilder}, \citenamefont {Ergun}, \citenamefont {Schwartz}, \citenamefont {Cassak}, \citenamefont {Burch}, \citenamefont {Chen}, \citenamefont {Torbert}, \citenamefont {Phan}, \citenamefont {Lavraud}, \citenamefont {Goodrich}, \citenamefont {Holmes}, \citenamefont {Stawarz}, \citenamefont {Sturner}, \citenamefont {Malaspina}, \citenamefont {Usanova}, \citenamefont {Trattner}, \citenamefont {Strangeway}, \citenamefont {Russell}, \citenamefont {Pollock}, \citenamefont {Giles}, \citenamefont {Hesse}, \citenamefont {Lindqvist}, \citenamefont {Drake}, \citenamefont {Shay}, \citenamefont {Nakamura},\ and\ \citenamefont {Marklund}}]{Eriksson2016}%
  \BibitemOpen
  \bibfield  {author} {\bibinfo {author} {\bibfnamefont {S.}~\bibnamefont {Eriksson}}, \bibinfo {author} {\bibfnamefont {F.~D.}\ \bibnamefont {Wilder}}, \bibinfo {author} {\bibfnamefont {R.~E.}\ \bibnamefont {Ergun}}, \bibinfo {author} {\bibfnamefont {S.~J.}\ \bibnamefont {Schwartz}}, \bibinfo {author} {\bibfnamefont {P.~A.}\ \bibnamefont {Cassak}}, \bibinfo {author} {\bibfnamefont {J.~L.}\ \bibnamefont {Burch}}, \bibinfo {author} {\bibfnamefont {L.-J.}\ \bibnamefont {Chen}}, \bibinfo {author} {\bibfnamefont {R.~B.}\ \bibnamefont {Torbert}}, \bibinfo {author} {\bibfnamefont {T.~D.}\ \bibnamefont {Phan}}, \bibinfo {author} {\bibfnamefont {B.}~\bibnamefont {Lavraud}}, \bibinfo {author} {\bibfnamefont {K.~A.}\ \bibnamefont {Goodrich}}, \bibinfo {author} {\bibfnamefont {J.~C.}\ \bibnamefont {Holmes}}, \bibinfo {author} {\bibfnamefont {J.~E.}\ \bibnamefont {Stawarz}}, \bibinfo {author} {\bibfnamefont {A.~P.}\ \bibnamefont {Sturner}}, \bibinfo {author} {\bibfnamefont {D.~M.}\ \bibnamefont {Malaspina}}, \bibinfo
  {author} {\bibfnamefont {M.~E.}\ \bibnamefont {Usanova}}, \bibinfo {author} {\bibfnamefont {K.~J.}\ \bibnamefont {Trattner}}, \bibinfo {author} {\bibfnamefont {R.~J.}\ \bibnamefont {Strangeway}}, \bibinfo {author} {\bibfnamefont {C.~T.}\ \bibnamefont {Russell}}, \bibinfo {author} {\bibfnamefont {C.~J.}\ \bibnamefont {Pollock}}, \bibinfo {author} {\bibfnamefont {B.~L.}\ \bibnamefont {Giles}}, \bibinfo {author} {\bibfnamefont {M.}~\bibnamefont {Hesse}}, \bibinfo {author} {\bibfnamefont {P.-A.}\ \bibnamefont {Lindqvist}}, \bibinfo {author} {\bibfnamefont {J.~F.}\ \bibnamefont {Drake}}, \bibinfo {author} {\bibfnamefont {M.~A.}\ \bibnamefont {Shay}}, \bibinfo {author} {\bibfnamefont {R.}~\bibnamefont {Nakamura}}, \ and\ \bibinfo {author} {\bibfnamefont {G.~T.}\ \bibnamefont {Marklund}},\ }\href {\doibase 10.1103/PhysRevLett.117.015001} {\bibfield  {journal} {\bibinfo  {journal} {Physical Review Letters}\ }\textbf {\bibinfo {volume} {117}},\ \bibinfo {pages} {015001} (\bibinfo {year} {2016})}\BibitemShut
  {NoStop}%
\bibitem [{\citenamefont {Ji}\ \emph {et~al.}(2023)\citenamefont {Ji}, \citenamefont {Yoo}, \citenamefont {Fox}, \citenamefont {Yamada}, \citenamefont {Argall}, \citenamefont {Egedal}, \citenamefont {Liu}, \citenamefont {Wilder}, \citenamefont {Eriksson}, \citenamefont {Daughton}, \citenamefont {Bergstedt}, \citenamefont {Bose}, \citenamefont {Burch}, \citenamefont {Torbert}, \citenamefont {Ng},\ and\ \citenamefont {Chen}}]{Ji2023pub}%
  \BibitemOpen
  \bibfield  {author} {\bibinfo {author} {\bibfnamefont {H.}~\bibnamefont {Ji}}, \bibinfo {author} {\bibfnamefont {J.}~\bibnamefont {Yoo}}, \bibinfo {author} {\bibfnamefont {W.}~\bibnamefont {Fox}}, \bibinfo {author} {\bibfnamefont {M.}~\bibnamefont {Yamada}}, \bibinfo {author} {\bibfnamefont {M.}~\bibnamefont {Argall}}, \bibinfo {author} {\bibfnamefont {J.}~\bibnamefont {Egedal}}, \bibinfo {author} {\bibfnamefont {Y.-H.}\ \bibnamefont {Liu}}, \bibinfo {author} {\bibfnamefont {R.}~\bibnamefont {Wilder}}, \bibinfo {author} {\bibfnamefont {S.}~\bibnamefont {Eriksson}}, \bibinfo {author} {\bibfnamefont {W.}~\bibnamefont {Daughton}}, \bibinfo {author} {\bibfnamefont {K.}~\bibnamefont {Bergstedt}}, \bibinfo {author} {\bibfnamefont {S.}~\bibnamefont {Bose}}, \bibinfo {author} {\bibfnamefont {J.}~\bibnamefont {Burch}}, \bibinfo {author} {\bibfnamefont {R.}~\bibnamefont {Torbert}}, \bibinfo {author} {\bibfnamefont {J.}~\bibnamefont {Ng}}, \ and\ \bibinfo {author} {\bibfnamefont {L.-J.}\ \bibnamefont {Chen}},\ }\href
  {\doibase 10.1007/s11214-023-01024-3} {\bibfield  {journal} {\bibinfo  {journal} {Space Science Reviews}\ }\textbf {\bibinfo {volume} {219}},\ \bibinfo {pages} {76} (\bibinfo {year} {2023})}\BibitemShut {NoStop}%
\bibitem [{\citenamefont {Frank}\ \emph {et~al.}(2006)\citenamefont {Frank}, \citenamefont {Bogdanov}, \citenamefont {Dreiden}, \citenamefont {Markov},\ and\ \citenamefont {Ostrovskaya}}]{Frank2006}%
  \BibitemOpen
  \bibfield  {author} {\bibinfo {author} {\bibfnamefont {A.~G.}\ \bibnamefont {Frank}}, \bibinfo {author} {\bibfnamefont {S.~{\relax Yu}.}\ \bibnamefont {Bogdanov}}, \bibinfo {author} {\bibfnamefont {G.~V.}\ \bibnamefont {Dreiden}}, \bibinfo {author} {\bibfnamefont {V.~S.}\ \bibnamefont {Markov}}, \ and\ \bibinfo {author} {\bibfnamefont {G.~V.}\ \bibnamefont {Ostrovskaya}},\ }\href {\doibase 10.1016/j.physleta.2005.08.049} {\bibfield  {journal} {\bibinfo  {journal} {Physics Letters A}\ }\textbf {\bibinfo {volume} {348}},\ \bibinfo {pages} {318} (\bibinfo {year} {2006})}\BibitemShut {NoStop}%
\bibitem [{\citenamefont {Bogdanov}\ \emph {et~al.}(2007)\citenamefont {Bogdanov}, \citenamefont {Dreiden}, \citenamefont {Markov}, \citenamefont {Ostrovskaya},\ and\ \citenamefont {Frank}}]{Bogdanov2007}%
  \BibitemOpen
  \bibfield  {author} {\bibinfo {author} {\bibfnamefont {S.~{\relax Yu}.}\ \bibnamefont {Bogdanov}}, \bibinfo {author} {\bibfnamefont {G.~V.}\ \bibnamefont {Dreiden}}, \bibinfo {author} {\bibfnamefont {V.~S.}\ \bibnamefont {Markov}}, \bibinfo {author} {\bibfnamefont {G.~V.}\ \bibnamefont {Ostrovskaya}}, \ and\ \bibinfo {author} {\bibfnamefont {A.~G.}\ \bibnamefont {Frank}},\ }\href {\doibase 10.1134/S1063780X07110050} {\bibfield  {journal} {\bibinfo  {journal} {Plasma Physics Reports}\ }\textbf {\bibinfo {volume} {33}},\ \bibinfo {pages} {930} (\bibinfo {year} {2007})}\BibitemShut {NoStop}%
\bibitem [{\citenamefont {Tharp}\ \emph {et~al.}(2012)\citenamefont {Tharp}, \citenamefont {Yamada}, \citenamefont {Ji}, \citenamefont {Lawrence}, \citenamefont {Dorfman}, \citenamefont {Myers},\ and\ \citenamefont {Yoo}}]{Tharp2012}%
  \BibitemOpen
  \bibfield  {author} {\bibinfo {author} {\bibfnamefont {T.~D.}\ \bibnamefont {Tharp}}, \bibinfo {author} {\bibfnamefont {M.}~\bibnamefont {Yamada}}, \bibinfo {author} {\bibfnamefont {H.}~\bibnamefont {Ji}}, \bibinfo {author} {\bibfnamefont {E.}~\bibnamefont {Lawrence}}, \bibinfo {author} {\bibfnamefont {S.}~\bibnamefont {Dorfman}}, \bibinfo {author} {\bibfnamefont {C.~E.}\ \bibnamefont {Myers}}, \ and\ \bibinfo {author} {\bibfnamefont {J.}~\bibnamefont {Yoo}},\ }\href {\doibase 10.1103/PhysRevLett.109.165002} {\bibfield  {journal} {\bibinfo  {journal} {Physical Review Letters}\ }\textbf {\bibinfo {volume} {109}},\ \bibinfo {pages} {165002} (\bibinfo {year} {2012})}\BibitemShut {NoStop}%
\bibitem [{\citenamefont {Fox}\ \emph {et~al.}(2017)\citenamefont {Fox}, \citenamefont {Park}, \citenamefont {Deng}, \citenamefont {Fiksel}, \citenamefont {Spitkovsky},\ and\ \citenamefont {Bhattacharjee}}]{Fox2017}%
  \BibitemOpen
  \bibfield  {author} {\bibinfo {author} {\bibfnamefont {W.}~\bibnamefont {Fox}}, \bibinfo {author} {\bibfnamefont {J.}~\bibnamefont {Park}}, \bibinfo {author} {\bibfnamefont {W.}~\bibnamefont {Deng}}, \bibinfo {author} {\bibfnamefont {G.}~\bibnamefont {Fiksel}}, \bibinfo {author} {\bibfnamefont {A.}~\bibnamefont {Spitkovsky}}, \ and\ \bibinfo {author} {\bibfnamefont {A.}~\bibnamefont {Bhattacharjee}},\ }\href {\doibase 10.1063/1.4993204} {\bibfield  {journal} {\bibinfo  {journal} {Physics of Plasmas}\ }\textbf {\bibinfo {volume} {24}},\ \bibinfo {pages} {092901} (\bibinfo {year} {2017})}\BibitemShut {NoStop}%
\bibitem [{\citenamefont {Hare}\ \emph {et~al.}(2018)\citenamefont {Hare}, \citenamefont {Suttle}, \citenamefont {Lebedev}, \citenamefont {Loureiro}, \citenamefont {Ciardi}, \citenamefont {Chittenden}, \citenamefont {Clayson}, \citenamefont {Eardley}, \citenamefont {Garcia}, \citenamefont {Halliday}, \citenamefont {Robinson}, \citenamefont {Smith}, \citenamefont {Stuart}, \citenamefont {{Suzuki-Vidal}},\ and\ \citenamefont {Tubman}}]{Hare2018a}%
  \BibitemOpen
  \bibfield  {author} {\bibinfo {author} {\bibfnamefont {J.~D.}\ \bibnamefont {Hare}}, \bibinfo {author} {\bibfnamefont {L.~G.}\ \bibnamefont {Suttle}}, \bibinfo {author} {\bibfnamefont {S.~V.}\ \bibnamefont {Lebedev}}, \bibinfo {author} {\bibfnamefont {N.~F.}\ \bibnamefont {Loureiro}}, \bibinfo {author} {\bibfnamefont {A.}~\bibnamefont {Ciardi}}, \bibinfo {author} {\bibfnamefont {J.}~\bibnamefont {Chittenden}}, \bibinfo {author} {\bibfnamefont {T.}~\bibnamefont {Clayson}}, \bibinfo {author} {\bibfnamefont {S.~J.}\ \bibnamefont {Eardley}}, \bibinfo {author} {\bibfnamefont {C.}~\bibnamefont {Garcia}}, \bibinfo {author} {\bibfnamefont {J.~W.~D.}\ \bibnamefont {Halliday}}, \bibinfo {author} {\bibfnamefont {T.}~\bibnamefont {Robinson}}, \bibinfo {author} {\bibfnamefont {R.~A.}\ \bibnamefont {Smith}}, \bibinfo {author} {\bibfnamefont {N.}~\bibnamefont {Stuart}}, \bibinfo {author} {\bibfnamefont {F.}~\bibnamefont {{Suzuki-Vidal}}}, \ and\ \bibinfo {author} {\bibfnamefont {E.~R.}\ \bibnamefont {Tubman}},\ }\href
  {\doibase 10.1063/1.5016280} {\bibfield  {journal} {\bibinfo  {journal} {Physics of Plasmas}\ }\textbf {\bibinfo {volume} {25}},\ \bibinfo {pages} {055703} (\bibinfo {year} {2018})}\BibitemShut {NoStop}%
\bibitem [{\citenamefont {Suttle}\ \emph {et~al.}(2016)\citenamefont {Suttle}, \citenamefont {Hare}, \citenamefont {Lebedev}, \citenamefont {Swadling}, \citenamefont {Burdiak}, \citenamefont {Ciardi}, \citenamefont {Chittenden}, \citenamefont {Loureiro}, \citenamefont {Niasse}, \citenamefont {{Suzuki-Vidal}}, \citenamefont {Wu}, \citenamefont {Yang}, \citenamefont {Clayson}, \citenamefont {Frank}, \citenamefont {Robinson}, \citenamefont {Smith},\ and\ \citenamefont {Stuart}}]{Suttle2016}%
  \BibitemOpen
  \bibfield  {author} {\bibinfo {author} {\bibfnamefont {L.~G.}\ \bibnamefont {Suttle}}, \bibinfo {author} {\bibfnamefont {J.~D.}\ \bibnamefont {Hare}}, \bibinfo {author} {\bibfnamefont {S.~V.}\ \bibnamefont {Lebedev}}, \bibinfo {author} {\bibfnamefont {G.~F.}\ \bibnamefont {Swadling}}, \bibinfo {author} {\bibfnamefont {G.~C.}\ \bibnamefont {Burdiak}}, \bibinfo {author} {\bibfnamefont {A.}~\bibnamefont {Ciardi}}, \bibinfo {author} {\bibfnamefont {J.~P.}\ \bibnamefont {Chittenden}}, \bibinfo {author} {\bibfnamefont {N.~F.}\ \bibnamefont {Loureiro}}, \bibinfo {author} {\bibfnamefont {N.}~\bibnamefont {Niasse}}, \bibinfo {author} {\bibfnamefont {F.}~\bibnamefont {{Suzuki-Vidal}}}, \bibinfo {author} {\bibfnamefont {J.}~\bibnamefont {Wu}}, \bibinfo {author} {\bibfnamefont {Q.}~\bibnamefont {Yang}}, \bibinfo {author} {\bibfnamefont {T.}~\bibnamefont {Clayson}}, \bibinfo {author} {\bibfnamefont {A.}~\bibnamefont {Frank}}, \bibinfo {author} {\bibfnamefont {T.~S.}\ \bibnamefont {Robinson}}, \bibinfo {author}
  {\bibfnamefont {R.~A.}\ \bibnamefont {Smith}}, \ and\ \bibinfo {author} {\bibfnamefont {N.}~\bibnamefont {Stuart}},\ }\href {\doibase 10.1103/PhysRevLett.116.225001} {\bibfield  {journal} {\bibinfo  {journal} {Physical Review Letters}\ }\textbf {\bibinfo {volume} {116}},\ \bibinfo {pages} {225001} (\bibinfo {year} {2016})}\BibitemShut {NoStop}%
\bibitem [{\citenamefont {Suttle}\ \emph {et~al.}(2018)\citenamefont {Suttle}, \citenamefont {Hare}, \citenamefont {Lebedev}, \citenamefont {Ciardi}, \citenamefont {Loureiro}, \citenamefont {Burdiak}, \citenamefont {Chittenden}, \citenamefont {Clayson}, \citenamefont {Halliday}, \citenamefont {Niasse}, \citenamefont {Russell}, \citenamefont {{Suzuki-Vidal}}, \citenamefont {Tubman}, \citenamefont {Lane}, \citenamefont {Ma}, \citenamefont {Robinson}, \citenamefont {Smith},\ and\ \citenamefont {Stuart}}]{Suttle2018}%
  \BibitemOpen
  \bibfield  {author} {\bibinfo {author} {\bibfnamefont {L.~G.}\ \bibnamefont {Suttle}}, \bibinfo {author} {\bibfnamefont {J.~D.}\ \bibnamefont {Hare}}, \bibinfo {author} {\bibfnamefont {S.~V.}\ \bibnamefont {Lebedev}}, \bibinfo {author} {\bibfnamefont {A.}~\bibnamefont {Ciardi}}, \bibinfo {author} {\bibfnamefont {N.~F.}\ \bibnamefont {Loureiro}}, \bibinfo {author} {\bibfnamefont {G.~C.}\ \bibnamefont {Burdiak}}, \bibinfo {author} {\bibfnamefont {J.~P.}\ \bibnamefont {Chittenden}}, \bibinfo {author} {\bibfnamefont {T.}~\bibnamefont {Clayson}}, \bibinfo {author} {\bibfnamefont {J.~W.~D.}\ \bibnamefont {Halliday}}, \bibinfo {author} {\bibfnamefont {N.}~\bibnamefont {Niasse}}, \bibinfo {author} {\bibfnamefont {D.}~\bibnamefont {Russell}}, \bibinfo {author} {\bibfnamefont {F.}~\bibnamefont {{Suzuki-Vidal}}}, \bibinfo {author} {\bibfnamefont {E.}~\bibnamefont {Tubman}}, \bibinfo {author} {\bibfnamefont {T.}~\bibnamefont {Lane}}, \bibinfo {author} {\bibfnamefont {J.}~\bibnamefont {Ma}}, \bibinfo {author}
  {\bibfnamefont {T.}~\bibnamefont {Robinson}}, \bibinfo {author} {\bibfnamefont {R.~A.}\ \bibnamefont {Smith}}, \ and\ \bibinfo {author} {\bibfnamefont {N.}~\bibnamefont {Stuart}},\ }\href {\doibase 10.1063/1.5023664} {\bibfield  {journal} {\bibinfo  {journal} {Physics of Plasmas}\ }\textbf {\bibinfo {volume} {25}},\ \bibinfo {pages} {042108} (\bibinfo {year} {2018})}\BibitemShut {NoStop}%
\bibitem [{\citenamefont {Hare}\ \emph {et~al.}(2017{\natexlab{b}})\citenamefont {Hare}, \citenamefont {Suttle}, \citenamefont {Lebedev}, \citenamefont {Loureiro}, \citenamefont {Ciardi}, \citenamefont {Burdiak}, \citenamefont {Chittenden}, \citenamefont {Clayson}, \citenamefont {Garcia}, \citenamefont {Niasse}, \citenamefont {Robinson}, \citenamefont {Smith}, \citenamefont {Stuart}, \citenamefont {Swadling}, \citenamefont {Ma}, \citenamefont {Wu},\ and\ \citenamefont {Yang}}]{Hare2017}%
  \BibitemOpen
  \bibfield  {author} {\bibinfo {author} {\bibfnamefont {J.~D.}\ \bibnamefont {Hare}}, \bibinfo {author} {\bibfnamefont {L.}~\bibnamefont {Suttle}}, \bibinfo {author} {\bibfnamefont {S.~V.}\ \bibnamefont {Lebedev}}, \bibinfo {author} {\bibfnamefont {N.~F.}\ \bibnamefont {Loureiro}}, \bibinfo {author} {\bibfnamefont {A.}~\bibnamefont {Ciardi}}, \bibinfo {author} {\bibfnamefont {G.~C.}\ \bibnamefont {Burdiak}}, \bibinfo {author} {\bibfnamefont {J.~P.}\ \bibnamefont {Chittenden}}, \bibinfo {author} {\bibfnamefont {T.}~\bibnamefont {Clayson}}, \bibinfo {author} {\bibfnamefont {C.}~\bibnamefont {Garcia}}, \bibinfo {author} {\bibfnamefont {N.}~\bibnamefont {Niasse}}, \bibinfo {author} {\bibfnamefont {T.}~\bibnamefont {Robinson}}, \bibinfo {author} {\bibfnamefont {R.~A.}\ \bibnamefont {Smith}}, \bibinfo {author} {\bibfnamefont {N.}~\bibnamefont {Stuart}}, \bibinfo {author} {\bibfnamefont {G.~F.}\ \bibnamefont {Swadling}}, \bibinfo {author} {\bibfnamefont {J.}~\bibnamefont {Ma}}, \bibinfo {author} {\bibfnamefont
  {J.}~\bibnamefont {Wu}}, \ and\ \bibinfo {author} {\bibfnamefont {Q.}~\bibnamefont {Yang}},\ }\href {\doibase 10.1103/PhysRevLett.118.085001} {\bibfield  {journal} {\bibinfo  {journal} {Physical Review Letters}\ }\textbf {\bibinfo {volume} {118}},\ \bibinfo {pages} {085001} (\bibinfo {year} {2017}{\natexlab{b}})}\BibitemShut {NoStop}%
\bibitem [{\citenamefont {Shay}, \citenamefont {Drake},\ and\ \citenamefont {Rogers}(2001)}]{Shay2001}%
  \BibitemOpen
  \bibfield  {author} {\bibinfo {author} {\bibfnamefont {M.~A.}\ \bibnamefont {Shay}}, \bibinfo {author} {\bibfnamefont {J.~F.}\ \bibnamefont {Drake}}, \ and\ \bibinfo {author} {\bibfnamefont {B.~N.}\ \bibnamefont {Rogers}},\ }\href {\doibase 10.1029/1999JA001007} {\bibfield  {journal} {\bibinfo  {journal} {Journal of Geophysical Research}\ }\textbf {\bibinfo {volume} {106}},\ \bibinfo {pages} {3759} (\bibinfo {year} {2001})}\BibitemShut {NoStop}%
\bibitem [{\citenamefont {Bola{\~n}os}\ \emph {et~al.}(2022)\citenamefont {Bola{\~n}os}, \citenamefont {Sladkov}, \citenamefont {Smets}, \citenamefont {Chen}, \citenamefont {Grisollet}, \citenamefont {Filippov}, \citenamefont {Henares}, \citenamefont {Nastasa}, \citenamefont {Pikuz}, \citenamefont {Riquier}, \citenamefont {Safronova}, \citenamefont {Severin}, \citenamefont {Starodubtsev},\ and\ \citenamefont {Fuchs}}]{Bolanos2022}%
  \BibitemOpen
  \bibfield  {author} {\bibinfo {author} {\bibfnamefont {S.}~\bibnamefont {Bola{\~n}os}}, \bibinfo {author} {\bibfnamefont {A.}~\bibnamefont {Sladkov}}, \bibinfo {author} {\bibfnamefont {R.}~\bibnamefont {Smets}}, \bibinfo {author} {\bibfnamefont {S.~N.}\ \bibnamefont {Chen}}, \bibinfo {author} {\bibfnamefont {A.}~\bibnamefont {Grisollet}}, \bibinfo {author} {\bibfnamefont {E.}~\bibnamefont {Filippov}}, \bibinfo {author} {\bibfnamefont {J.-L.}\ \bibnamefont {Henares}}, \bibinfo {author} {\bibfnamefont {V.}~\bibnamefont {Nastasa}}, \bibinfo {author} {\bibfnamefont {S.}~\bibnamefont {Pikuz}}, \bibinfo {author} {\bibfnamefont {R.}~\bibnamefont {Riquier}}, \bibinfo {author} {\bibfnamefont {M.}~\bibnamefont {Safronova}}, \bibinfo {author} {\bibfnamefont {A.}~\bibnamefont {Severin}}, \bibinfo {author} {\bibfnamefont {M.}~\bibnamefont {Starodubtsev}}, \ and\ \bibinfo {author} {\bibfnamefont {J.}~\bibnamefont {Fuchs}},\ }\href {\doibase 10.1038/s41467-022-33813-9} {\bibfield  {journal} {\bibinfo  {journal} {Nature
  Communications}\ }\textbf {\bibinfo {volume} {13}},\ \bibinfo {pages} {6426} (\bibinfo {year} {2022})}\BibitemShut {NoStop}%
\bibitem [{\citenamefont {Swadling}\ \emph {et~al.}(2014)\citenamefont {Swadling}, \citenamefont {Lebedev}, \citenamefont {Hall}, \citenamefont {Patankar}, \citenamefont {Stewart}, \citenamefont {Smith}, \citenamefont {{Harvey-Thompson}}, \citenamefont {Burdiak}, \citenamefont {{de Grouchy}}, \citenamefont {Skidmore}, \citenamefont {Suttle}, \citenamefont {{Suzuki-Vidal}}, \citenamefont {Bland}, \citenamefont {Kwek}, \citenamefont {Pickworth}, \citenamefont {Bennett}, \citenamefont {Hare}, \citenamefont {Rozmus},\ and\ \citenamefont {Yuan}}]{Swadling2014a}%
  \BibitemOpen
  \bibfield  {author} {\bibinfo {author} {\bibfnamefont {G.~F.}\ \bibnamefont {Swadling}}, \bibinfo {author} {\bibfnamefont {S.~V.}\ \bibnamefont {Lebedev}}, \bibinfo {author} {\bibfnamefont {G.~N.}\ \bibnamefont {Hall}}, \bibinfo {author} {\bibfnamefont {S.}~\bibnamefont {Patankar}}, \bibinfo {author} {\bibfnamefont {N.~H.}\ \bibnamefont {Stewart}}, \bibinfo {author} {\bibfnamefont {R.~A.}\ \bibnamefont {Smith}}, \bibinfo {author} {\bibfnamefont {A.~J.}\ \bibnamefont {{Harvey-Thompson}}}, \bibinfo {author} {\bibfnamefont {G.~C.}\ \bibnamefont {Burdiak}}, \bibinfo {author} {\bibfnamefont {P.}~\bibnamefont {{de Grouchy}}}, \bibinfo {author} {\bibfnamefont {J.}~\bibnamefont {Skidmore}}, \bibinfo {author} {\bibfnamefont {L.}~\bibnamefont {Suttle}}, \bibinfo {author} {\bibfnamefont {F.}~\bibnamefont {{Suzuki-Vidal}}}, \bibinfo {author} {\bibfnamefont {S.~N.}\ \bibnamefont {Bland}}, \bibinfo {author} {\bibfnamefont {K.~H.}\ \bibnamefont {Kwek}}, \bibinfo {author} {\bibfnamefont {L.}~\bibnamefont {Pickworth}},
  \bibinfo {author} {\bibfnamefont {M.}~\bibnamefont {Bennett}}, \bibinfo {author} {\bibfnamefont {J.~D.}\ \bibnamefont {Hare}}, \bibinfo {author} {\bibfnamefont {W.}~\bibnamefont {Rozmus}}, \ and\ \bibinfo {author} {\bibfnamefont {J.}~\bibnamefont {Yuan}},\ }\href {\doibase 10.1063/1.4890564} {\bibfield  {journal} {\bibinfo  {journal} {Review of Scientific Instruments}\ }\textbf {\bibinfo {volume} {85}},\ \bibinfo {pages} {11E502} (\bibinfo {year} {2014})}\BibitemShut {NoStop}%
\bibitem [{\citenamefont {Hare}\ \emph {et~al.}(2019)\citenamefont {Hare}, \citenamefont {MacDonald}, \citenamefont {Bland}, \citenamefont {Dranczewski}, \citenamefont {Halliday}, \citenamefont {Lebedev}, \citenamefont {Suttle}, \citenamefont {Tubman},\ and\ \citenamefont {Rozmus}}]{Hare2019}%
  \BibitemOpen
  \bibfield  {author} {\bibinfo {author} {\bibfnamefont {J.~D.}\ \bibnamefont {Hare}}, \bibinfo {author} {\bibfnamefont {J.}~\bibnamefont {MacDonald}}, \bibinfo {author} {\bibfnamefont {S.~N.}\ \bibnamefont {Bland}}, \bibinfo {author} {\bibfnamefont {J.}~\bibnamefont {Dranczewski}}, \bibinfo {author} {\bibfnamefont {J.~W.~D.}\ \bibnamefont {Halliday}}, \bibinfo {author} {\bibfnamefont {S.~V.}\ \bibnamefont {Lebedev}}, \bibinfo {author} {\bibfnamefont {L.~G.}\ \bibnamefont {Suttle}}, \bibinfo {author} {\bibfnamefont {E.~R.}\ \bibnamefont {Tubman}}, \ and\ \bibinfo {author} {\bibfnamefont {W.}~\bibnamefont {Rozmus}},\ }\href {\doibase 10.1088/1361-6587/ab2571} {\bibfield  {journal} {\bibinfo  {journal} {Plasma Physics and Controlled Fusion}\ }\textbf {\bibinfo {volume} {61}},\ \bibinfo {pages} {085012} (\bibinfo {year} {2019})}\BibitemShut {NoStop}%
\bibitem [{\citenamefont {Datta}\ \emph {et~al.}(2022)\citenamefont {Datta}, \citenamefont {Russell}, \citenamefont {Tang}, \citenamefont {Clayson}, \citenamefont {Suttle}, \citenamefont {Chittenden}, \citenamefont {Lebedev},\ and\ \citenamefont {Hare}}]{Datta2022a}%
  \BibitemOpen
  \bibfield  {author} {\bibinfo {author} {\bibfnamefont {R.}~\bibnamefont {Datta}}, \bibinfo {author} {\bibfnamefont {D.~R.}\ \bibnamefont {Russell}}, \bibinfo {author} {\bibfnamefont {I.}~\bibnamefont {Tang}}, \bibinfo {author} {\bibfnamefont {T.}~\bibnamefont {Clayson}}, \bibinfo {author} {\bibfnamefont {L.~G.}\ \bibnamefont {Suttle}}, \bibinfo {author} {\bibfnamefont {J.~P.}\ \bibnamefont {Chittenden}}, \bibinfo {author} {\bibfnamefont {S.~V.}\ \bibnamefont {Lebedev}}, \ and\ \bibinfo {author} {\bibfnamefont {J.~D.}\ \bibnamefont {Hare}},\ }\href {\doibase 10.1063/5.0098823} {\bibfield  {journal} {\bibinfo  {journal} {Review of Scientific Instruments}\ }\textbf {\bibinfo {volume} {93}},\ \bibinfo {pages} {103530} (\bibinfo {year} {2022})}\BibitemShut {NoStop}%
\bibitem [{\citenamefont {Lebedev}\ \emph {et~al.}(2001)\citenamefont {Lebedev}, \citenamefont {Beg}, \citenamefont {Bland}, \citenamefont {Chittenden}, \citenamefont {Dangor}, \citenamefont {Haines}, \citenamefont {Kwek}, \citenamefont {Pikuz},\ and\ \citenamefont {Shelkovenko}}]{Lebedev2001}%
  \BibitemOpen
  \bibfield  {author} {\bibinfo {author} {\bibfnamefont {S.~V.}\ \bibnamefont {Lebedev}}, \bibinfo {author} {\bibfnamefont {F.~N.}\ \bibnamefont {Beg}}, \bibinfo {author} {\bibfnamefont {S.~N.}\ \bibnamefont {Bland}}, \bibinfo {author} {\bibfnamefont {J.~P.}\ \bibnamefont {Chittenden}}, \bibinfo {author} {\bibfnamefont {A.~E.}\ \bibnamefont {Dangor}}, \bibinfo {author} {\bibfnamefont {M.~G.}\ \bibnamefont {Haines}}, \bibinfo {author} {\bibfnamefont {K.~H.}\ \bibnamefont {Kwek}}, \bibinfo {author} {\bibfnamefont {S.~A.}\ \bibnamefont {Pikuz}}, \ and\ \bibinfo {author} {\bibfnamefont {T.~A.}\ \bibnamefont {Shelkovenko}},\ }\href {\doibase 10.1063/1.1385373} {\bibfield  {journal} {\bibinfo  {journal} {Physics of Plasmas}\ }\textbf {\bibinfo {volume} {8}},\ \bibinfo {pages} {3734} (\bibinfo {year} {2001})}\BibitemShut {NoStop}%
\bibitem [{\citenamefont {Chittenden}\ \emph {et~al.}(2004)\citenamefont {Chittenden}, \citenamefont {Lebedev}, \citenamefont {Oliver}, \citenamefont {Yu},\ and\ \citenamefont {Cuneo}}]{Chittenden2004}%
  \BibitemOpen
  \bibfield  {author} {\bibinfo {author} {\bibfnamefont {J.~P.}\ \bibnamefont {Chittenden}}, \bibinfo {author} {\bibfnamefont {S.~V.}\ \bibnamefont {Lebedev}}, \bibinfo {author} {\bibfnamefont {B.~V.}\ \bibnamefont {Oliver}}, \bibinfo {author} {\bibfnamefont {E.~P.}\ \bibnamefont {Yu}}, \ and\ \bibinfo {author} {\bibfnamefont {M.~E.}\ \bibnamefont {Cuneo}},\ }\href {\doibase 10.1063/1.1643756} {\bibfield  {journal} {\bibinfo  {journal} {Physics of Plasmas}\ }\textbf {\bibinfo {volume} {11}},\ \bibinfo {pages} {1118} (\bibinfo {year} {2004})}\BibitemShut {NoStop}%
\bibitem [{\citenamefont {Ciardi}\ \emph {et~al.}(2007)\citenamefont {Ciardi}, \citenamefont {Lebedev}, \citenamefont {Frank}, \citenamefont {Blackman}, \citenamefont {Chittenden}, \citenamefont {Jennings}, \citenamefont {Ampleford}, \citenamefont {Bland}, \citenamefont {Bott}, \citenamefont {Rapley}, \citenamefont {Hall}, \citenamefont {{Suzuki-Vidal}}, \citenamefont {Marocchino}, \citenamefont {Lery},\ and\ \citenamefont {Stehle}}]{Ciardi2007}%
  \BibitemOpen
  \bibfield  {author} {\bibinfo {author} {\bibfnamefont {A.}~\bibnamefont {Ciardi}}, \bibinfo {author} {\bibfnamefont {S.~V.}\ \bibnamefont {Lebedev}}, \bibinfo {author} {\bibfnamefont {A.}~\bibnamefont {Frank}}, \bibinfo {author} {\bibfnamefont {E.~G.}\ \bibnamefont {Blackman}}, \bibinfo {author} {\bibfnamefont {J.~P.}\ \bibnamefont {Chittenden}}, \bibinfo {author} {\bibfnamefont {C.~J.}\ \bibnamefont {Jennings}}, \bibinfo {author} {\bibfnamefont {D.~J.}\ \bibnamefont {Ampleford}}, \bibinfo {author} {\bibfnamefont {S.~N.}\ \bibnamefont {Bland}}, \bibinfo {author} {\bibfnamefont {S.~C.}\ \bibnamefont {Bott}}, \bibinfo {author} {\bibfnamefont {J.}~\bibnamefont {Rapley}}, \bibinfo {author} {\bibfnamefont {G.~N.}\ \bibnamefont {Hall}}, \bibinfo {author} {\bibfnamefont {F.~A.}\ \bibnamefont {{Suzuki-Vidal}}}, \bibinfo {author} {\bibfnamefont {A.}~\bibnamefont {Marocchino}}, \bibinfo {author} {\bibfnamefont {T.}~\bibnamefont {Lery}}, \ and\ \bibinfo {author} {\bibfnamefont {C.}~\bibnamefont {Stehle}},\ }\href
  {\doibase 10.1063/1.2436479} {\bibfield  {journal} {\bibinfo  {journal} {Physics of Plasmas}\ }\textbf {\bibinfo {volume} {14}} (\bibinfo {year} {2007}),\ 10.1063/1.2436479},\ \Eprint {http://arxiv.org/abs/astro-ph/0611441} {arXiv:astro-ph/0611441} \BibitemShut {NoStop}%
\end{thebibliography}%

\end{document}